\documentclass[english,twocolumn,transaction]{IEEEtran}
\usepackage[T1]{fontenc}
\usepackage{float}
\usepackage{amsmath}
\usepackage{amsthm}
\usepackage{amssymb}
\usepackage{stackrel}
\usepackage{graphicx}
\usepackage{setspace}
\usepackage{color}
\usepackage{url}

\makeatletter

\providecommand{\tabularnewline}{\\}
\floatstyle{ruled}
\newfloat{algorithm}{tbp}{loa}
\providecommand{\algorithmname}{Algorithm}
\floatname{algorithm}{\protect\algorithmname}

\theoremstyle{plain}

\theoremstyle{definition}

\theoremstyle{plain}

\theoremstyle{plain}

\IEEEoverridecommandlockouts
\usepackage{ifpdf}
\usepackage{cite}
\ifCLASSOPTIONcompsoc
\usepackage[caption=false,font=normalsize,labelfont=sf,textfont=sf]{subfig}
\else
\usepackage[caption=false,font=footnotesize]{subfig}
\fi
\usepackage{amsmath}

\@ifundefined{showcaptionsetup}{}{%
 \PassOptionsToPackage{caption=false}{subfig}}
\usepackage{subfig}
\makeatother

\usepackage{babel}
\newtheorem{define}{Definition}
\newtheorem{coro}{Corollary}
\newtheorem{lemm}{Lemma}
\newtheorem{theo}{Theorem}
\newtheorem{remark}{Remark}
\renewcommand\figurename{Fig.}
\begin{document}

\title{Energy-Efficient Caching for Scalable Videos in Heterogeneous
Networks}

\author{Xuewei Zhang, ~\IEEEmembership{Student Member,~IEEE}, Tiejun Lv, ~\IEEEmembership{Senior Member,~IEEE}, \\
Wei Ni, ~\IEEEmembership{Senior Member,~IEEE}, John M. Cioffi, ~\IEEEmembership{Fellow,~IEEE}, \\
Norman C. Beaulieu, ~\IEEEmembership{Fellow,~IEEE}, and Y. Jay Guo, ~\IEEEmembership{Fellow,~IEEE}
\thanks{The financial support of the National Natural Science Foundation of
China (NSFC) (Grant No. 61671072) is gratefully acknowledged. (\emph{Corresponding author: Tiejun Lv.})

X. Zhang, T. Lv and N. C. Beaulieu are with the School of Information and Communication
Engineering, Beijing University of Posts and Telecommunications (BUPT), Beijing
100876, China (e-mail: \{zhangxw, lvtiejun, nborm\}@bupt.edu.cn).

W. Ni is with Data61, Commonwealth Scientific and Industrial Research (e-mail: wei.ni@data61.csiro.au).

J. M. Cioffi is with the Department of Electrical Engineering, Stanford
University, Stanford, CA 94305 USA (e-mail: cioffi@stanford.edu).

Y. J. Guo is with the Global Big Data Technologies Centre, University of Technology Sydney, Australia
(e-mail: jay.guo@uts.edu.au).
}}

\maketitle
\begin{abstract}
By suppressing repeated content deliveries,
wireless caching has the potential to substantially improve the energy efficiency (EE) of the fifth generation (5G) communication networks.
In this paper, we propose two novel energy-efficient caching schemes in heterogeneous networks,
namely, scalable video coding (SVC)-based fractional caching and SVC-based random caching,
which can provide on-demand video services with different perceptual qualities.
We derive the expressions for successful transmission probabilities and ergodic service rates.
Based on the derivations and the established power consumption models,
the EE maximization problems are formulated for the two proposed caching schemes.
By taking logarithmic approximations of the $l_{0}$-norm, the problems are efficiently solved by the standard gradient projection method.
Numerical results validate the theoretical analysis
and demonstrate the superiority of our proposed caching schemes, compared to three benchmark strategies.
\end{abstract}

\begin{IEEEkeywords}
Energy efficiency (EE), heterogeneous networks, scalable video coding (SVC), standard gradient projection method, wireless caching.
\end{IEEEkeywords}
\section{Introduction}
\renewcommand\figurename{Fig.}
Exposed to information explosion and data tsunami,
we have witnessed explosive traffic surges for socializing, working and entertainment.
It is forecasted that the total amount of data traffic is expected to achieve at 100 exabytes in 2023,
and over 75\% of this traffic is expected to be generated from bandwidth-demanding multimedia video services \cite{2016Ericsson}.
Notably, there are a large number of repeated deliveries for popular video files \cite{wang2014cache},
which would cause huge resource wastes and aggravate traffic burdens over backhaul links.
Therefore, innovative techniques are desired to address the repeated deliveries of bandwidth-demanding popular videos.
In light of this, wireless caching has been proposed as the appealing candidate technique
in the fifth generation (5G) communication networks \cite{poularakis2016exploiting}.
Wireless caching can effectively relieve the severe traffic burden over backhaul links
and reduce service delay \cite{Tao2015Content,zhou2015content,Dong2016Cache}.
Additionally, it also exhibits strong potential to
reduce power consumption and improve energy efficiency (EE) of wireless systems \cite{Wu2017An}.

Wireless caching allows base stations (BSs) to prefetch video files
from the core network through capacity-limited backhaul links during off-peak hours.
These videos can be placed in the local storage of the BSs \cite{Chen2017Cooperative123},
and delivered to users when requested.
Thereby, it can relieve the requests of backhaul bandwidth during peak-hours.
Depending on different content placement strategies, caching can be typically classified into two categories,
including uncoded caching \cite{zhou2015content,Tao2015Content}
and coded caching \cite{Xu2017Fundamental,Liao2016Optimizing,Gabry2016On,Chen2015Optimal}.
Uncoded caching aims at storing complete video files in each of the BS,
which is very suitable for popular videos.
In order to effectively leverage the cumulative cache size of nearby BSs,
coded caching enables each BS to store different fragments or proportions of the encoded contents,
and a recipient of the video can construct the content file based on a set of pre-defined decoding rules.
Recently, the network coding-based caching has been accepted as a promising content placement scheme,
such as the maximum distance separable (MDS) coding-based caching schemes \cite{Liao2016Optimizing,Gabry2016On,Chen2015Optimal,Xu2017Modeling}.
Random caching schemes have also gained a lot of interest \cite{Wen2017Random,cui2016analysis},
where video files or their combinations are
randomly placed
under certain probability distributions to yield optimal successful transmission probability.
Note that random caching can be regarded as a special case of uncoded caching, since complete video files are cached;
however, the caching probabilities have yet to be determined.
Till now, wireless caching has been extensively studied in
cloud radio access networks (C-RAN) \cite{Tao2015Content,zhao2016cluster},
heterogeneous networks \cite{Gabry2016On,Wen2017Random},
device-to-device (D2D) communications \cite{chen2016cache,Zhang2015Efficient},
small cell networks (SCNs) \cite{tamoor2016caching,Hu2016Caching,Xu2017Modeling}
and networks combined with D2D and SCNs\cite{yang2016analysis}, etc.
Consensus has been reached that wireless caching is able to reduce power consumption,
service delay and backhaul-link traffic burden, as well as improve the request hit ratio.

Perceptual requirements of video subscribers can have strong impact on the design of wireless caching system.
For example, people generally request standard viewing quality for news reports and sports games,
and high viewing quality for movies and TV series.
However, this issue has not been captured in the
aforementioned existing studies.
Scalable video coding (SVC), as part of the H.265 standard \cite{Schwarz2007Overview},
is able to flexibly remove part of the video bit streams to adjust to various user requirements
and network states while guaranteeing acceptable video quality.
To elaborate a little further,
in SVC, each video file is divided into a base layer (BL) and multiple enhancement layers (ELs).
Videos with only BL can provide fundamental viewing quality,
while EL contents can complement to the BL to provide superior video quality.
It is worth noting that EL contents cannot be decoded without the corresponding BL\cite{Ostovari2015Scalable}.
Some research efforts have been devoted to combining SVC with wireless caching.
For example, Wu \emph{et al.} in \cite{Wu2016Caching} analyze the successful transmission rate
and backhaul load in cellular networks by caching and transmitting
scalable videos, and confirm the effectiveness of taking different viewing quality requirements into consideration.
Additionally, the authors in \cite{ZhanContent} propose an SVC-based layer placement strategy,
which can significantly reduce the average content download time.
These works focus on analyzing the
successful transmission rate, backhaul load and average content download time.
In the near future, more diverse and comprehensive performance metrics, such as EE,
are advocated so that more performance gain can be obtained through the combination of wireless caching and SVC.

In a different yet relevant context of 5G,
EE is an important design criterion \cite{Peng2015Energy,Zhao2017Joint},
where the energy saving should not be
at the cost of the quality of service (QoS) \cite{Wu2017An,Zhang2017Fundamental}.
The optimal EE design is critical,
and can make preferable balance between total power consumption and spectrum efficiency (SE)
so as to improve resource utilization \cite{Wang2017Energy}.
It is obvious that this issue also merits consideration in cache-enabled networks.
Specifically, the authors in \cite{Liu2015Energy} provide the closed-form expression for EE and identify the conditions to obtain benefits from caching.
Chen \emph{et al.} in \cite{Chen2017Cooperative123} and Zhang \emph{et al.} in \cite{zhang2016cache} also derive the expressions for EE
by exploiting stochastic geometry.
Furthermore, some researches focus on the design of optimal content placement policy to enhance EE.
To be more specific, the authors in \cite{Gabry2016On} minimize the total power consumption to improve EE
by designing the placement scheme of the coded packets under the MDS coding-based caching strategy.
These works have been focusing on the optimization of EE,
especially the minimization of total power consumption to improve EE.
They fail to apply to the more practical scenario,
where different viewing requirements need to be taken into consideration.
It is of great importance to provide scalable video services
when concentrating on EE optimizations in cache-enabled networks.

To fill this void, in this paper, we investigate the energy-efficient caching schemes
to yield optimal EE performance in cache-enabled heterogeneous networks,
where video files to be requested are encoded by SVC
and are divided into a BL and an EL.
The BL and EL contents are locally cached and cooperatively transmitted by two clusters of small-cell BSs (SBSs).
We propose two caching schemes,
namely, SVC-based fractional caching and SVC-based random caching, to improve the EE of the cache-enabled networks.
A comprehensive performance analysis is carried out to characterize the EE of cache-enabled networks supporting SVC.
To derive the closed-form expressions for EE, we establish the power consumption models and theoretically analyze the ergodic service rates.
Afterwards, the optimization problems are formulated to maximize EE of the two proposed caching schemes,
which are then effectively solved by the standard gradient projection method after taking the approximations of $l_{0}$-norm.

The main contributions of this paper are summarized as follows:
\begin{itemize}
\item We propose two new SVC-based caching schemes to support scalable video services.
Both the content popularity and the quality preference of the videos are taken into account.
In specific, standard definition video (SDV) and high definition video (HDV) can both be provided to the users,
which are able to improve resource utilization and and thereby enhance EE.
\item Relying on stochastic geometry,
when the designated user is served by the nearest MBS and cooperative SBSs,
the closed-form expressions for successful transmission probabilities and ergodic service rates are derived,
from which some useful insights are shed.
\item It is more challenging to deal with the proposed EE optimization problems.
After taking logarithmic approximations of the $l_{0}$-norm,
we transform the objective functions into the continuous and differential ones.
Finally, according to their specific forms,
the EE optimization problems are efficiently solved by the proposed standard gradient projection method.
\end{itemize}

$\;$The rest of this paper is organized as follows.
Section II presents the system model,
including the network model, SVC-based caching schemes,
channel model and power consumption model of the cache-enabled heterogeneous networks.
Closed-form expressions for successful transmission probabilities
and ergodic service rates are derived in Section III.
Under the two proposed SVC-based caching schemes,
the EE optimization problems are formulated and solved in Section IV and Section V, respectively.
Extensive simulation results are presented in Section VI,
and this paper concludes in Section VII.
\section{System Model}
In this section, we present the system model of the cache-enabled networks supporting SVC,
including the network model, SVC-based caching schemes,
channel model and power consumption model.
\subsection{Network Model}
Fig. 1 illustrates the heterogeneous network of interest, including an MBS tier and an SBS tier.
Multiple MBSs and SBSs are independently and identically distributed in the network of interest,
whose locations follow homogeneous Poisson Point Processes (PPPs) $\mathrm{\Phi_{M}}$
and $\mathrm{\Phi_{S}}$ with densities $\lambda_{M}$ and $\lambda _{S}$, respectively.
Without loss of generality, our analysis is carried out for a designated user,
which is located at the center of the network
\cite{Wen2017Random,zhang2016cache,Chen2017Cooperative123,Huang2014Energy}.
All the BSs and the user are assumed to be equipped with a single antenna.
To implement different caching and transmission assignments, the SBSs are grouped into two clusters.
In specific, taking the user's position as the center, SBSs within the circle with radius $a$ form cluster $\mathcal{N}_{1}$
and the number of SBSs in this cluster is $N_{1}=\left|\mathcal{N}_{1}\right|$.
The SBSs located in the annulus with radii $a$ and $b$ ($a<b$) form cluster $\mathcal{N}_{2}$,
and the number of SBSs in this cluster is $N_{2}=\left|\mathcal{N}_{2}\right|$.

Employing SVC, each video file is encoded into a BL and an EL
\footnote{In this paper, we consider two-layer video caching and transmission for illustration convenience.
When the video files are divided into more layers,
there can be more clusters of SBSs to cooperatively serve the user,
while the performance analysis and the problem formulation can follow the exactly same steps proposed in the paper.
To this end, the proposed two-layer video caching and transmission model is instrumental,
and can be readily extended to the multi-layer cases.}.
The BL provides fundamental video quality,
while users with both BL and EL contents can acquire superior video quality.
Video files with only BL contents are defined as SDVs,
and videos with both BL and EL contents are defined as HDVs \cite{Wu2016Caching}.
The SBSs in cluster $\mathcal{N}_{1}$ are assigned to cache BL contents,
while ELs are cached in the local storage of SBSs belonging to cluster $\mathcal{N}_{2}$
\footnote{BL content is the most fundamental part of a scalable video,
since the decoding of EL largely depends on the received BL.
Owing to the paramount importance of BLs, the SBSs located in cluster $\mathcal{N}_{1}$,
which are closer to the user and are capable of providing stronger signal strength than those in cluster $\mathcal{N}_{2}$,
are responsible for caching and delivering BLs, while SBSs located in cluster $\mathcal{N}_{2}$ are assigned to provide ELs.}.
When the designated user requests the $f$-th video file,
the SBSs located in cluster $\mathcal{N}_{1}$ are responsible for transmitting the BL content of this file;
and the SBSs located in cluster $\mathcal{N}_{2}$ will deliver the EL content if HDV is required.
Some SBSs in clusters $\mathcal{N}_{1}$ and $\mathcal{N}_{2}$ that do not possess the required BL and EL contents remain silent
until the next transmission process begins.
When cooperative SBSs fail to provide the complete video layers,
the nearest MBS is activated to deliver the remaining video contents from the core networks via backhaul links.

\begin{figure}[t]
\centering{}\includegraphics[scale=0.26]{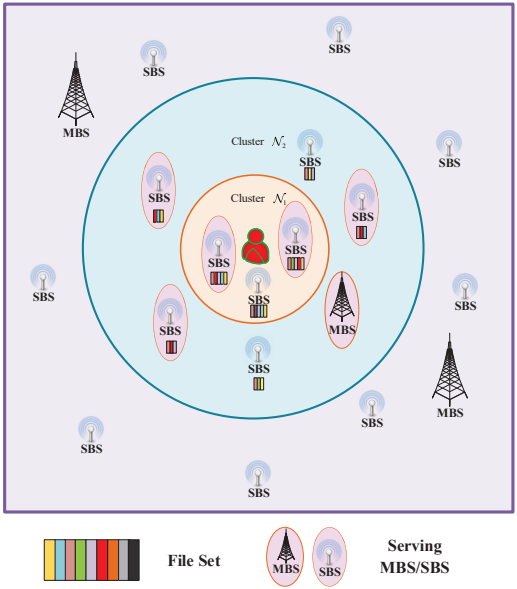}
\caption{In the proposed system model,
the designated user is located at the center of the observed network
and requests the video file colored in red.
The serving SBSs are those who cache the BL and EL contents colored in red and are located in the circle and annular areas.
If the user cannot obtain complete video layers,
the nearest MBS will be triggered to retrieve the remaining video contents.}
\end{figure}
\subsection{SVC-Based Caching Schemes}
Assume that there are a total of $F$ video files requested by the user.
The sizes of each BL and EL are $L_{B}$ and $L_{E}$,
respectively\footnote{Without loss of generality, equal video sizes are assumed in this paper.
Of course, we can also consider different content sizes in the proposed schemes,
which has little effect on the performance analysis.}.
The total cache size at each SBS is denoted by $M$.
According to what mentioned above,
the number of locally cached BL and EL contents can be given by $M_{B}=\left\lfloor (M/L_{B})\right\rfloor$
and $M_{E}=\left\lfloor (M/L_{E})\right\rfloor $.
All videos are arranged in the descending order of popularity,
where more popular videos are ranked with smaller indices.
We assume that the probability of video requests follows the Zipf's law, as given by \cite{shanmugam2013femtocaching}
\begin{equation}
p_{f}=\frac{f^{-\alpha}}{\sum_{n=1}^{F}n^{-\alpha}},\ f=1,2,...,F,\label{eq:c-1-1}
\end{equation}
\noindent where $\alpha$ is the skewness parameter
that characterizes the concentration of video requests \cite{breslau1999web}.
The viewing quality preference for SDV and HDV is also considered.
According to \cite{Wu2016Caching},
the SDV perceptual preference of the $f$-th video file can be modeled as
\begin{equation}
g_{\rm{SDV}}(f)=\frac{f-1}{F-1},\ f=1,2,...,F.
\end{equation}
\noindent The HDV perceptual preference can be accordingly denoted as $g_{\rm{HDV}}(f)=1-g_{\rm{SDV}}(f)$.

Based on the above content transmission protocol and user request model,
we propose two caching schemes, which are described as follows:
\begin{itemize}
\item Scheme I (SVC-based fractional caching).
The SBSs located in cluster $\mathcal{N}_{1}$ cache parts of the BL contents of the video files,
where the cached parts are the same across the SBSs.
For SBSs located in cluster $\mathcal{N}_{2}$,
some parts of the EL contents are cached in their local storage,
where the cached parts are also the same across these SBSs.
Let $Q_{1,f}$ and $Q_{2,f}$ denote the caching fractions of the BL and EL contents of the $f$-th video file in
SBSs belonging to clusters $\mathcal{N}_{1}$ and $\mathcal{N}_{2}$, respectively.
They can be stacked into vectors $\mathbf{Q}_{1}=[Q_{1,1},Q_{1,2},...,Q_{1,F}]$ and $\mathbf{Q}_{2}=[Q_{2,1},Q_{2,2},...,Q_{2,F}]$.
\item Scheme II (SVC-based random caching). The SBSs located in clusters
$\mathcal{N}_{1}$ and $\mathcal{N}_{2}$ randomly cache the complete BL and EL contents under certain probability distributions, respectively.
Let $T_{1,f}$ and $T_{2,f}$
denote the probabilities for caching BL and EL contents of the $f$-th
video file in SBSs belonging to clusters $\mathcal{N}_{1}$ and $\mathcal{N}_{2}$, respectively.
They can be stacked into vectors $\mathbf{T}_{1}=[T_{1,1},T_{1,2},...,T_{1,F}]$ and $\mathbf{T}_{2}=[T_{2,1},T_{2,2},...,T_{2,F}]$.
\end{itemize}

In the following sections,
we will derive the optimal caching distributions to yield optimal EE of the cache-enabled network.

\subsection{Channel Model}
Interference typically dominates over noises in modern cellular networks.
This paper is therefore focusing on an interference-limited case,
where the background additive white Gaussian noise is comparatively negligible at the user, as compared to the co-channel interference \cite{Wen2017Random,Wu2016Caching,Xu2017Modeling}.
When the cached BL content of the $f$-th video file is received,
the signal-to-interference ratio (SIR) of the designated user is expressed as
\begin{align}
&\mathrm{SIR}_{S,BL}=\nonumber\\
&\frac{\left|\sum_{k\in\mathcal{N}_{1,f}}h_{S,k}\sqrt{P_{S}}r_{S,k}^{-\frac{\alpha_{S}}{2}}\right|^{2}}{\sum_{n\in\mathrm{\Phi_{S}}\setminus\mathcal{N}_{1}}\left|h_{S,n}\right|^{2}P_{S}r_{S,n}^{-\alpha_{S}}+\sum_{m\in\mathrm{\Phi_{M}}}\left|h_{M,m}\right|^{2}P_{M}r_{M,m}^{-\alpha_{M}}},\label{eq:SINR_BL_S}
\end{align}
\noindent where $\mathcal{N}_{1,f}$ denotes
the set of SBSs located in cluster $\mathcal{N}_{1}$ that cache the BL content of the $f$-th video file.
$h_{S,k}$ and $h_{M,m}$ are the channel gains from the $k$-th SBS and the $m$-th MBS,
following the complex Gaussian distribution
with zero mean and unit variance, i.e., $\mathcal{CN}\sim(0,1)$.
$P_{S}$ and $P_{M}$ are the transmit powers of SBS and MBS, respectively.
$r_{S,k}$ is the distance between the $k$-th SBS and the user.
Likewise, $r_{M,m}$ is the distance between the $m$-th MBS and the user.
Additionally, $\alpha_{S}$ and $\alpha_{M}$ are the path loss exponents from the SBSs and MBSs to the user.
In (\ref{eq:SINR_BL_S}), the first term in the denominator of the right hand side accounts for the interference from other SBSs,
except those located in cluster $\mathcal{N}_{1}$,
and the second term represents the cross-tier interference from all the MBSs.
When the cached EL content of the $f$-th video file is delivered to the user,
the received SIR is given by
\begin{align}
&\mathrm{SIR}_{S,EL}=\nonumber\\
&\frac{\left|\sum_{k\in\mathcal{N}_{2,f}}h_{S,k}\sqrt{P_{S}}r_{S,k}^{-\frac{\alpha_{S}}{2}}\right|^{2}}{\sum_{n\in\mathrm{\Phi_{S}}\setminus\mathcal{N}_{2}}\left|h_{S,n}\right|^{2}P_{S}r_{S,n}^{-\alpha_{S}}+\sum_{m\in\mathrm{\Phi_{M}}}\left|h_{M,m}\right|^{2}P_{M}r_{M,m}^{-\alpha_{M}}},
\end{align}
\noindent where $\mathcal{N}_{2,f}$ denotes the set of SBSs located in cluster $\mathcal{N}_{2}$
that cache EL content of the $f$-th video file.
As mentioned, the required video layers may not be completely provided by cooperative SBSs.
In this case, the remaining video contents can proceed to be delivered by the nearest MBS.
As a result, the received SIR is denoted as
\begin{align}
&\mathrm{SIR}_{M}=\nonumber\\
&\frac{\left|h_{M,m_{0}}\right|^{2}P_{M}r_{M,m_{0}}^{-\alpha_{M}}}{\sum_{n\in\mathrm{\Phi_{S}}}\left|h_{S,n}\right|^{2}P_{S}r_{S,n}^{-\alpha_{S}}+\sum_{m\in\mathrm{\Phi_{M}\setminus}m_{0}}\left|h_{M,m}\right|^{2}P_{M}r_{M,m}^{-\alpha_{M}}},\label{eq:SINR_M}
\end{align}
\noindent where $m_{0}$ refers to the nearest MBS that provides the user with the required video content.
\subsection{Power Consumption Model}
The total power consumption of the cache-enabled network can be modeled as
\begin{gather}
P_{\rm{Total}}=P_{\rm{TR}}+P_{\rm{CA}}+P_{\rm{BH}}+P_{\rm{Fix}},\label{eq:power_1}
\end{gather}
where $P_{\rm{TR}}$, $P_{\rm{CA}}$, $P_{\rm{BH}}$ and $P_{\rm{Fix}}$ are the power consumptions for data transmission,
content caching, backhaul delivery and other fixed budgets, respectively.
Details of these kinds of  power consumptions are illustrated as follows.

After placing parts of the video layers in the local storage of SBSs,
the caching power consumption exists.
As described in \cite{Gabry2016On,Xu2014Coordinated},
the caching power consumption is proportional to the number of the data
bits stored in the BSs. Therefore, the caching power consumption is calculated as
\begin{equation}
P_{\rm{CA}}=c_{ca}N_{ca},
\end{equation}
\noindent where $c_{ca}$ is the caching coefficient in $\mathrm{W/bit}$
and $N_{ca}$ is the total number of data bits
cached in the local storage of SBSs belonging to clusters $\mathcal{N}_{1}$ and $\mathcal{N}_{2}$.

Due to the limited storage of SBSs,
all required video layers cannot be locally cached.
Parts of them need to be retrieved from the nearest MBS via backhaul links.
This leads to the backhaul power consumption.
The backhaul consumption is proportional to the total number of the data bits transmitted via backhaul links,
which is given by
\begin{align}
 P_{\rm{BH}}=c_{bh}N_{bh},\label{eq:vuy}
\end{align}
\noindent where $c_{bh}$ is the coefficient of backhaul power consumption in $\mathrm{W/bit}$
and $N_{bh}$ is the total number of data bits delivered by backhaul links.
In practical implementations, for video files with the same sizes,
caching them typically consumes less power consumption than delivering them via microwave backhaul links.
In this sense, $c_{ca}$ is typically less than $c_{bh}$ \cite{Liu2015Energy}.

As for the fixed power consumption $P_{\rm{Fix}}$,
it is mainly caused by site-cooling, controlling and running circuit components and the oscillator.
In the proposed network model, the fixed power consumption is calculated as
\begin{gather}
P_{\rm{Fix}}=(N_{1}+N_{2})P_{\rm{S,Fix}}+P_{\rm{M,Fix}},\label{eq:vufg}
\end{gather}
\noindent where $P_{\rm{S,Fix}}$ and $P_{\rm{M,Fix}}$ are the
fixed power consumption constants for the SBSs and MBSs, respectively.
\section{The Ergodic Service Rate}
In this section, we derive the expressions for successful transmission probabilities and ergodic service rates
when the user is served by the nearest MBS and cooperative SBSs,
which provides the corner stone to derive the expressions for EE.
\subsection{The Ergodic Service Rate From the Nearest MBS}
Given limited storage capacity of SBSs,
they can hardly cache all of the BL and EL contents locally.
When the complete content layers of the required videos cannot be delivered to the user's side,
the nearest MBS proceeds to provide the remaining video contents.
In the following,
we derive the ergodic service rate in the case where the user is served by its nearest MBS.
\begin{define}
When the nearest MBS provides the user with the required video layers,
the ergodic service rate is written as
\begin{align}
R_{M}(\gamma)\triangleq W\mathrm{\mathbb{E}}\left\{ \log_{2}(1+\mathrm{SIR}_{M})|\mathrm{SIR}_{M}\geq\gamma\right\} ,\label{eq:Service Rate_M}
\end{align}
\noindent where $W$ is the spectrum bandwidth and $\gamma$
is set as the minimum QoS requirement.
$\mathbb{E}\left\{\cdot\right\}$ takes the expectation with respect to small-scale fading,
and the locations of PPP-distributed MBSs and SBSs.
\end{define}
Note that in (\ref{eq:Service Rate_M}),
the minimum QoS requirement is guaranteed with the given $R_{M}(\gamma)$.
To derive the expression for $R_{M}(\gamma)$, we have to first derive the successful transmission
probability, i.e., $P(\mathrm{SIR}_{M}\geq\gamma)$, which can be provided by Lemma 1.
\begin{lemm}
When the user is served by the nearest MBS under the minimum QoS requirement $\gamma$,
the successful transmission probability is given in (\ref{eq:P_SINR_M}), where $G_{\alpha}(x)=\int_{x}^{\infty}\frac{1}{1+t^{\frac{\alpha}{2}}}\mathrm{d}t$.
\begin{algorithm*}[t]
\begin{align}
& P(\mathrm{SIR}_{M}\geq\gamma)=2\pi\lambda_{M}\int_{0}^{\infty}x\exp(-\pi(\lambda_{S}(\gamma\frac{P_{S}}{P_{M}}x^{\alpha_{M}})^{\frac{2}{\alpha_{S}}}G_{\alpha_{S}}(0)-\lambda_{M}x^{2}(\gamma^{\frac{2}{\alpha_{M}}}G_{\alpha_{M}}(\gamma{}^{-\frac{2}{\alpha_{M}}})+1)))\mathrm{d}x,\label{eq:P_SINR_M}
\end{align}
\begin{align}
P(\mathrm{SIR}_{S,BL}\geq\gamma_{BL})=&\int_{0}^{a}\int_{0}^{a}...\int_{0}^{a}\stackrel[k=1]{n_{1}}{\prod}\frac{2x_{BL,k}}{a^{2}}
\exp(-\pi(\lambda_{S}c{}^{\frac{2}{\alpha_{S}}}G_{\alpha_{S}}(a^{2}c^{-\frac{2}{\alpha_{S}}})-\lambda_{M}(c\frac{P_{M}}{P_{S}})^{\frac{2}{\alpha_{M}}}G_{\alpha_{M}}(0)))\mathrm{d}\mathbf{x}_{BL},\label{eq:P_S_BL}
\end{align}
\begin{align}
P(\mathrm{SIR}_{S,EL}\geq\gamma_{EL})=&\int_{a}^{b}\int_{a}^{b}...\int_{a}^{b}\stackrel[k=1]{n_{2}}{\prod}\frac{2x_{EL,k}}{b^{2}-a^{2}}\nonumber \\
 &\exp(-\pi(\lambda_{M}(d\frac{P_{M}}{P_{S}})^{\frac{2}{\alpha_{M}}}G_{\alpha_{M}}(0)-\lambda_{S}d{}^{\frac{2}{\alpha_{S}}}(\int_{0}^{a^{2}d^{-\frac{2}{\alpha_{S}}}}\frac{1}{1+t^{\frac{\alpha_{S}}{2}}}\mathrm{d}t+G_{\alpha_{S}}(b^{2}d^{-\frac{2}{\alpha_{S}}}))))\mathrm{d}\mathbf{x}_{EL},\label{eq:P_S_EL}
\end{align}
\end{algorithm*}
\end{lemm}

\begin{IEEEproof}
See Appendix A.
\end{IEEEproof}

From (\ref{eq:P_SINR_M}), we can find that the derived expression for $P(\mathrm{SIR}_{M}\geq\gamma)$ can be complicated.
Fortunately, when $\alpha_{M}=\alpha_{S}=4$,
(\ref{eq:P_SINR_M}) can be expressed in a much simpler form, as shown in the following corollary.
\begin{coro}
When $\alpha_{M}=\alpha_{S}=4$, the expression for $P(\mathrm{SIR}_{M}\geq\gamma)$ can be simplified as
\begin{align}
&P(\mathrm{SIR}_{M}\geq\gamma)=\nonumber\\
&(1+\lambda_{M}^{-1}\gamma{}^{\frac{1}{2}}(\frac{\pi}{2}\lambda_{S}(\frac{P_{S}}{P_{M}})^{\frac{1}{2}}+\lambda_{M}\mathrm{arccot}(\gamma{}^{-\frac{1}{2}})))^{-1}.\label{eq:P_SINR_M_Special}
\end{align}
\end{coro}
\begin{IEEEproof}
See Appendix B.
\end{IEEEproof}

In Corollary 1, the simplified form of $P(\mathrm{SIR}_{M}\geq\gamma)$ is obtained,
from which some useful insights are shed, as follows.
\begin{itemize}
\item With the increasing ratio between the transmit powers of SBS and MBS,
i.e., $\frac{P_{S}}{P_{M}}$, $P(\mathrm{SIR}_{M}\geq\gamma)$ degrades.
This is due to the fact that SBSs with larger signal strength can produce stronger cross-tier interference
while the useful signal strength remains the same.
\item With the increasing ratio between the PPP densities of SBS and MBS, i.e., $\frac{\lambda_{S}}{\lambda_{M}}$,
$P(\mathrm{SIR}_{M}\geq\gamma)$ deteriorates.
This is because there are more SBSs, leading to severer cross-tier interference.
\item As the minimum QoS requirement $\gamma$ increases,
the user can impose strict video quality requirement
and $P(\mathrm{SIR}_{M}\geq\gamma)$ decreases.
The reason is because the co-exiting intra-tier and inter-tier interference restricts
further improvement of the successful transmission probability.
\end{itemize}

Based on the expression for $P(\mathrm{SIR}_{M}\geq\gamma)$ in Lemma 1,
$R_{M}(\gamma)$ can be calculated by employing the following theorem.
\begin{theo}
When the user is served by the nearest MBS, the ergodic service rate is expressed as
\begin{align}\label{theo1}
 &R_{M}(\gamma)=W\log_{2}(1+\gamma)+\frac{2\pi\lambda_{M}W}{\ln2}\nonumber\\
 &\int_{0}^{\infty}xe^{-\lambda_{M}\pi x^{2}}\mathrm{d}x\int_{\gamma}^{\infty}\frac{P(\mathrm{SIR}_{M}\geq t|r_{M,m_{0}}=x)}{P(\mathrm{SIR}_{M}\geq\gamma|r_{M,m_{0}}=x)(1+t)}\mathrm{d}t,
\end{align}
\noindent where the expressions for $P(\mathrm{SIR}_{M}\geq t|r_{M,m_{0}}=x)$
and $P(\mathrm{SIR}_{M}\geq\gamma|r_{M,m_{0}}=x)$ can be found in (\ref{eq:vbyu}), as given in Appendix A.
\end{theo}
\begin{IEEEproof}
See Appendix C.
\end{IEEEproof}
\begin{remark}
From Theorem 1,
it is obvious that a many of factors have impact on the ergodic service rate,
such as the minimum QoS requirement, the PPP densities of SBSs and MBSs, and the system bandwidth.
Among them, the minimum QoS requirement plays the most important role in the ergodic service rate,
since the two terms shown in (\ref{theo1}) are both related to the minimum QoS requirement.
As the QoS requirement grows, the first term of (\ref{theo1}) increases.
In the meanwhile, the successful transmission probability decreases,
leading to a reduction of the second term of (\ref{theo1}).
When the increase of the first term can compensate for the loss of the second term,
the ergodic service rate is enhanced.
Otherwise, the performance deteriorates.
Thus, it is not always the case that high QoS requirement would lead to good ergodic service rate.
\end{remark}
\subsection{The Ergodic Service Rates From Cooperative SBSs}
We proceed to derive the expressions for ergodic service rates
when the required BL and EL contents are delivered by cooperative SBSs,
which are denoted by $R_{S,BL}$ and $R_{S,EL}$, respectively.
The definitions of $R_{S,BL}$ and $R_{S,EL}$ are given in the following.

\begin{define}
When cooperative SBSs provide the user with the required BL and EL contents,
the ergodic service rates are defined as
\begin{gather}
 R_{S,BL}(\gamma_{BL},n_{1})\triangleq W\mathrm{\mathbb{E}}\left\{ \log_{2}(1+\mathrm{SIR}_{S,BL})|\mathrm{SIR}_{S,BL}\geq\gamma_{BL}\right\}, \label{eq:Rate_S_BL}
\end{gather}
\begin{gather}
 R_{S,EL}(\gamma_{EL},n_{2})\triangleq W\mathrm{\mathbb{E}}\left\{ \log_{2}(1+\mathrm{SIR}_{S,EL})|\mathrm{SIR}_{S,EL}\geq\gamma_{EL}\right\},\label{Rate_M_EL}
\end{gather}
\noindent where $\gamma_{BL}$ and $\gamma_{EL}$ are the minimum QoS requirements
for delivering BL and EL contents, respectively.
$n_{1}=\left|\mathcal{N}_{1,f}\right|\leq N_{1}$
and $n_{2}=\left|\mathcal{N}_{2,f}\right|\leq N_{2}$ are the numbers of the serving SBSs
that cache the required BL and EL contents.
\end{define}

\begin{remark}
According to Definitions 1 and 2,
it can be seen that the ergodic service rates are obtained
under the constraint of minimum QoS requirements.
Thus, in the EE optimization problems formulated later,
the QoS constraints are inherently satisfied in the devised EE expressions,
and we only focus on the cache size restrictions.
\end{remark}

In order to obtain the expressions devised in Definition 2,
we need to acquire the successful transmission probabilities
when the designated user receives BL and EL contents from cooperative SBSs,
i.e., $P(\mathrm{SIR}_{S,BL}\geq\gamma_{BL})$
and $P(\mathrm{SIR}_{S,EL}\geq\gamma_{EL})$. The successful transmission probabilities are given in the following lemma.

\begin{lemm}
When there are $n_{1}$ and $n_{2}$ SBSs to provide BL and EL contents, respectively,
the successful transmission probabilities are shown in (\ref{eq:P_S_BL}) and (\ref{eq:P_S_EL}),
where $c=\frac{\gamma_{BL}}{\sum_{k=1}^{n_{1}}x_{BL,k}^{-\alpha_{S}}}$, $d=\frac{\gamma_{EL}}{\sum_{k=1}^{n_{2}}x_{EL,k}^{-\alpha_{S}}}$,
$\mathbf{x}_{BL}=[x_{BL,1},...,x_{BL,n_{1}}]$
and $\mathbf{x}_{EL}=[x_{EL,1},...,x_{EL,n_{2}}]$.
\end{lemm}

\begin{IEEEproof}
See Appendix D.
\end{IEEEproof}

Considering the special case of $\alpha_{M}=\alpha_{S}=4$,
we are able to achieve the simplified forms of $P(\mathrm{SIR}_{S,BL}\geq\gamma_{BL})$
and $P(\mathrm{SIR}_{S,EL}\geq\gamma_{EL})$, as dictated in the following corollary.
\begin{coro}
When $\alpha_{M}=\alpha_{S}=4$, $P(\mathrm{SIR}_{S,BL}\geq\gamma_{BL})$
and $P(\mathrm{SIR}_{S,EL}\geq\gamma_{EL})$ can be simplified as
\begin{align}
&P(\mathrm{SIR}_{S,BL}\geq\gamma_{BL})=\nonumber\\
&\int_{0}^{a}...\int_{0}^{a}\stackrel[k=1]{n_{1}}{\prod}\frac{2x_{BL,k}}{a^{2}}\exp(-\pi uc{}^{\frac{1}{2}})\mathrm{d}\mathbf{x}_{BL},
\end{align}
\begin{align}
&P(\mathrm{SIR}_{S,EL}\geq\gamma_{EL})=\nonumber\\
&\int_{a}^{b}...\int_{a}^{b}\stackrel[k=1]{n_{2}}{\prod}\frac{2x_{EL,k}}{b^{2}-a^{2}}\exp(-\pi vd^{\frac{1}{2}})\mathrm{d}\mathbf{x}_{EL},
\end{align}
\noindent where
\begin{align}
u=\lambda_{S}\mathrm{arccot}(a^{2}c^{-\frac{1}{2}})+\frac{\pi}{2}\lambda_{M}(\frac{P_{M}}{P_{S}})^{\frac{1}{2}},
\end{align}
\begin{align}
v=\lambda_{S}(\arctan(a^{2}d^{-\frac{1}{2}})+\mathrm{arccot}(b^{2}d^{-\frac{1}{2}}))+\frac{\pi}{2}\lambda_{M}(\frac{P_{M}}{P_{S}})^{\frac{1}{2}}.
\end{align}
\end{coro}

\begin{IEEEproof}
Corollary 2 can be proved by following the steps shown in Appendix B, and therefore suppressed.
\end{IEEEproof}

Based on Lemma 2,
$R_{S,BL}(\gamma_{BL},n_{1})$ and $R_{S,EL}(\gamma_{EL},n_{2})$ can be derived, as presented in Theorem 2.

\begin{theo}
Let vectors $\mathbf{r}_{S_{BL}}=[r_{BL,1},r_{BL,2},...,r_{BL,n_{1}}]$ and $\mathbf{r}_{S_{EL}}=[r_{EL,1},r_{EL,2},...,r_{EL,n_{2}}]$
denote the positions of the serving SBSs belonging to clusters $\mathcal{N}_{1}$ and $\mathcal{N}_{2}$, respectively.
When the designated user is served by cooperative SBSs to obtain BL and EL contents,
the ergodic service rates can be calculated as
\begin{align}
 &R_{S,BL}(\gamma_{BL,}n_{1})=W\log_{2}(1+\gamma_{BL})\nonumber\\
 &+\frac{W}{\ln2}\int_{0}^{a}\int_{0}^{a}...\int_{0}^{a}\stackrel[k=1]{n_{1}}{\prod}\frac{2x_{BL,k}}{a^{2}}\mathrm{d}\mathbf{x}_{BL}\nonumber\\ &\int_{\gamma_{BL}}^{\infty}\frac{P(\mathrm{SIR}_{S,BL}\geq
 t|\mathbf{r}_{S_{BL}}=\mathbf{x}_{BL})}{P(\mathrm{SIR}_{S,BL}\geq\gamma_{BL}|\mathbf{r}_{S_{BL}}=\mathbf{x}_{BL})(1+t)}\mathrm{d}t, \label{eq:Rate_S_BL_1}
\end{align}
\begin{align}
 &R_{S,EL}(\gamma_{EL,}n_{2})=W\log_{2}(1+\gamma_{EL})\nonumber \\
 &+\frac{W}{\ln2}\int_{a}^{b}\int_{a}^{b}...\int_{a}^{b}\stackrel[k=1]{n_{2}}{\prod}\frac{2x_{EL,k}}{(b^{2}-a^{2})}\mathrm{d}\mathbf{x}_{EL}\nonumber\\
 &\int_{\gamma_{EL}}^{\infty}\frac{P(\mathrm{SIR}_{S,EL}\geq t|\mathbf{r}_{S_{EL}}=\mathbf{x}_{EL})}{P(\mathrm{SIR}_{S,EL}\geq\gamma_{EL}|\mathbf{r}_{S_{EL}}=\mathbf{x}_{EL})(1+t)}\mathrm{d}t,\label{eq:Rate_S_EL_1}
 \end{align}
\noindent where the expressions for $P(\mathrm{SIR}_{S,BL}\geq t|\mathbf{r}_{S_{BL}}=\mathbf{x}_{BL})$
and $P(\mathrm{SIR}_{S,BL}\geq\gamma_{BL}|\mathbf{r}_{S_{BL}}=\mathbf{x}_{BL})$
can be found in (\ref{eq:cv}), and the expressions for $P(\mathrm{SIR}_{S,EL}\geq t|\mathbf{r}_{S_{EL}}=\mathbf{x}_{EL})$
and $P(\mathrm{SIR}_{S,EL}\geq\gamma_{EL}|\mathbf{r}_{S_{EL}}=\mathbf{x}_{EL})$
can be found in (\ref{eq:bn}).
\end{theo}

\begin{IEEEproof}
This theorem can be proved by following the steps shown in Appendix C, and therefore omitted for brevity.
\end{IEEEproof}

From Theorem 2, some useful insights can also be observed,
which are similar to those shown in Remark 1.
For avoiding repetition, they are omitted here.
From the above derivations, we have successfully obtained the expressions for ergodic service rates,
which are the key intermediate steps to gain the EE expressions.
\section{The EE Optimization problem for Scheme I }
In this section, we derive the expressions for total power consumption and the sum rate,
based on which the EE maximization problem is formulated.
The proposed optimization problem is approximated, and then efficiently solved by the proposed standard gradient projection method.
\subsection{EE Optimization Problem Formulation}
Under Scheme I, the total power consumption of the network of interest can be quantified as
\begin{gather}
P_{\rm{Total,1}}=P_{\rm{TR,1}}+P_{\rm{CA,1}}+P_{\rm{BH,1}}+P_{\rm{Fix}},\label{eq:power_1}
\end{gather}
where
\begin{align}
P_{\rm{TR,1}}=&\sum_{f=1}^{F}p_{f}(\zeta_{S}(N_{1}\left\Vert Q_{1,f}\right\Vert _{0}+g_{\rm{HDV}}(f)N_{2}\left\Vert Q_{2,f}\right\Vert _{0})P_{S}\nonumber\\
&+\zeta_{M}(\left\Vert 1-Q_{1,f}\right\Vert _{0}+g_{\rm{HDV}}(f)\left\Vert 1-Q_{2,f}\right\Vert _{0})P_{M}),\label{TR_1}
\end{align}
\begin{equation}
P_{\rm{CA,1}}=c_{ca}\sum_{f=1}^{F}(Q_{1,f}L_{B}N_{1}+Q_{2,f}L_{E}N_{2}),
\end{equation}
\noindent and
\begin{align}
 P_{\rm{BH,1}}=c_{bh}\sum_{f=1}^{F}p_{f}((1-Q_{1,f})L_{B}+g_{\rm{HDV}}(1-Q_{2,f})L_{E}).\label{eq:vuy}
\end{align}
\noindent In (\ref{TR_1}), $\zeta_{S}$ and $\zeta_{M}$ are the power efficiency coefficients of the power amplifiers of the SBSs and MBSs, respectively,
and $\left\Vert \cdot\right\Vert _{0}$ stands for $l_{0}$-norm.

Based on the proposed SVC-based factional caching and cooperative transmission schemes,
the sum rate of the designated user is given by
\begin{align}
 & R_{\rm{Sum,1}}=\sum_{f=1}^{F}p_{f}\nonumber\\
 &(\underbrace{(1-Q_{1,f})R_{M}(\gamma_{BL})+g_{\rm{HDV}}(f)(1-Q_{2,f})R_{M}(\gamma_{EL})}_{\mathrm{served\,by\,the\,nearest\,MBS\,to\,obtain\,BL\,and\,EL\,contents}}\nonumber\\
 & +\underbrace{Q_{1,f}R_{S,BL}(\gamma_{BL},N_{1})+g_{\rm{HDV}}(f)Q_{2,f}R_{S,EL}(\gamma_{EL},N_{2})}_{\mathrm{served\:by\,cooperative\,SBSs\,to\,obtain\,BL\,and\,EL\,contents}}).\label{eq:R_SUM_1}
\end{align}
\noindent From (\ref{eq:R_SUM_1}),
intuitions can be found that the performance of $R_{\rm Sum,1}$ can
predominately depend on the caching fractions,
the number of serving SBSs and the minimum QoS requirements.
Moreover, if HDV service is required by the user,
the SBSs located in cluster $\mathcal{N}_{2}$ are activated to deliver the EL contents.
This can improve the sum rate performance.
According to the expressions for total power consumption and sum rate presented above,
the EE maximization problem with BL and EL caching fraction design
can be formulated as
\begin{subequations}\label{max_original}
\begin{align}
\underset{\mathbf{Q}_{1},\mathbf{Q}_{2}}{\mathrm{max}}\quad\: & EE_{1}=\frac{R_{\rm{Sum,1}}}{P_{\rm{Total,1}}}\label{eq:objective}\\
\mathrm{\mathrm{s.t.}}\:\quad\:\: & \sum_{f=1}^{F}Q_{1,f}=M_{B},\label{eq:CacheSizeConstraints_1}\\
 & \sum_{f=1}^{F}Q_{2,f}=M_{E},\label{eq:CacheSizeConstraints_2}\\
 & 0\leq Q_{1,f}\leq1,\,0\leq Q_{2,f}\leq1,\forall f,\label{eq:range}
\end{align}
\end{subequations}
\noindent where
(\ref{eq:CacheSizeConstraints_1}) and (\ref{eq:CacheSizeConstraints_2})
are the cache size restrictions of each SBS when caching BL and EL contents, respectively;
(\ref{eq:range}) indicates the feasible solution regions of the optimization variables $Q_{1,f}$ and $Q_{2,f}$.
\subsection{The Proposed Algorithm}
In (\ref{max_original}),
constraints (\ref{eq:CacheSizeConstraints_1}) to (\ref{eq:range}) form a convex variable set,
while the objective function has a complicated expression.
Moreover, the $l_{0}$-norm in $P_{\rm{Total},1}$ makes the problem more challenging to deal with.
Note that the $l_{0}$-norm can be approximated by a logarithmic function,
an exponential function or an arctangent function \cite{Tao2015Content}.
Without loss of generality, the logarithmic function is adopted in this paper as the smooth function.
Then, the $l_{0}$-norm in $P_{\rm{TR,1}}$ can be approximated as
\begin{align}
\hat{P}_{\rm{TR,1}}=&\sum_{f=1}^{F}p_{f}(\zeta_{S}(N_{1}f_{\theta}(Q_{1,f})+g_{\rm{HDV}}(f)N_{2}f_{\theta}(Q_{2,f}))P_{S}\nonumber\\
&+\zeta_{M}(f_{\theta}(1-Q_{1,f})+g_{\rm{HDV}}(f)f_{\theta}(1-Q_{2,f}))P_{M}),
\end{align}
\noindent where
\begin{align}
f_{\theta}(x)=\log(\frac{x}{\theta}+1)/\log(\frac{1}{\theta}+1).\label{SmoothFunction}
\end{align}
\noindent In \eqref{SmoothFunction}, $\theta$ is a constant parameter to reflect the smoothness of $f_{\theta}(x)$.
A larger value for $\theta$ leads to a smoother function but less accurate approximation.
%{\color{blue}After this transformation, (\ref{eq:objective}) is converted into a continuous and differentiable function of $Q_{1,f}$ and $Q_{2,f}$,
%but the convexity of the function is difficult to access.
%Though there are many EE optimization tools,
%such as those in \cite{Ng2013Wireless,Wu2016Energy,Miao2012Energy},
%they are not applicable in our proposed EE optimization problem.}
After the approximations, the objective function \eqref{eq:objective} is a continuous and differentiable function with respect to $Q_{1,f}$ and $Q_{2,f}$.
Due to the specific form of the transformed EE optimization problem,
the standard gradient projection algorithm is employed \cite{Wen2017Random,cui2016analysis,Bertsekas1997Nonlinear},
and the suboptimal solution of
(\ref{max_original}) can be obtained, as summarized in Algorithm 1.
In the proposed algorithm, $\epsilon(t)$ is the iteration step size,
which satisfies $\underset{t\rightarrow0}{\mathrm{lim}}\,\epsilon(t)=0,$
$\underset{T\rightarrow\infty}{\mathrm{lim}}\sum_{t=1}^{T}\epsilon(t)=\infty$ and
$\underset{T\rightarrow\infty}{\mathrm{lim}}\sum_{t=1}^{T}\epsilon^{2}(t)<\infty$.
We set $\epsilon(t)=\frac{1}{t}$ in the $t$-th iteration.
Note that Steps 3) and 5) give the projections of $\hat{Q}_{1,f}(t+1)$ and $\hat{Q}_{2,f}(t+1)$ onto the optimization variable set
so that constraints (\ref{eq:CacheSizeConstraints_1}) and (\ref{eq:CacheSizeConstraints_2}) are satisfied,
where $[x]^{+}\triangleq\rm{max}[x,0]$.

\begin{algorithm}[t]
\begin{enumerate}
\item Initialization: Set $t=1$, $\epsilon(1)=1$, and find $Q_{1,f}$ and
$Q_{2,f}$ that are feasible for constraints (\ref{eq:CacheSizeConstraints_1})-(\ref{eq:range}).
\item For $f\in\{1,...,F\}$, calculate $\frac{\partial EE_{1}}{Q_{1,f}}$,
and then obtain
\begin{align}
 \hat{Q}_{1,f}(t+1)=&Q_{1,f}(t)\nonumber\\
 &+\epsilon(t)\frac{\partial EE_{1}}{Q_{1,f}}\mid_{Q_{1,f}=Q_{1,f}(t),Q_{2,f}=Q_{2,f}(t).}\nonumber
\end{align}
\item For $f\in\{1,...,F\}$, calculate $Q_{1,f}(t+1)=\mathrm{min}\{[\hat{Q}_{1,f}(t+1)-u^{'}]^{+},1\}$,
where $u^{'}$ satisfies $\sum_{_{f=1}}^{F}\mathrm{min}\{[\hat{Q}_{1,f}(t+1)-u^{'}]^{+},1\}=M_{B}$.
\item For $f\in\{1,...,F\}$, calculate $\frac{\partial EE_{1}}{Q_{2,f}}$,
and then obtain
\begin{align}
 \hat{Q}_{2,f}(t+1)=&Q_{2,f}(t)\nonumber\\
 &+\epsilon(t)\frac{\partial EE_{1}}{Q_{2,f}}\mid_{Q_{1,f}=Q_{1,f}(t),Q_{2,f}=Q_{2,f}(t).}\nonumber
\end{align}
\item For $f\in\{1,...,F\}$, calculate $Q_{2,f}(t+1)=\mathrm{min}\{[\hat{Q}_{2,f}(t+1)-v^{'}]^{+},1\}$,
where $v^{'}$ satisfies $\sum_{_{f=1}}^{F}\mathrm{min}\{[\hat{Q}_{2,f}(t+1)-v^{'}]^{+},1\}=M_{E}$.
\item If convergence, the algorithm terminates. Otherwise, set $t=t+1$
and $\epsilon(t)=\frac{1}{t}$, then go back to Step 2).
\end{enumerate}
\caption{The proposed standard gradient projection method for solving the EE optimization problem (\ref{max_original}).}
\end{algorithm}
\section{The EE Optimization Problem for Scheme II}
For Scheme II, according to the power consumption model established previously,
the total power consumption is given by
\begin{equation}\label{pt_s2}
P_{\rm Total,2}=P_{\rm TR,2}+P_{\rm CA,2}+P_{\rm BH,2}+P_{\rm Fix},
\end{equation}
\noindent where
\begin{align}
P_{\rm{TR,2}}=&\sum_{f=1}^{F}p_{f}(\zeta_{S}(N_{1}T_{1,f}+g_{\rm{HDV}}(f)N_{2}T_{2,f})P_{S}\nonumber\\
&+\zeta_{M}((1-T_{1,f})+g_{\rm{HDV}}(f)(1-T_{2,f}))P_{M}),
\end{align}
\begin{align}
P_{\rm{CA,2}}=c_{ca}\sum_{f=1}^{F}(T_{1,f}L_{B}N_{1}+T_{2,f}L_{B}N_{2}),
\end{align}
\noindent and
\begin{gather}
 P_{\rm{BH,2}}=c_{bh}\sum_{f=1}^{F}p_{f}((1-T_{1,f})^{N_{1}}L_{B}+g_{\rm{HDV}}(1-T_{2,f})^{N_{2}}L_{E}).
\end{gather}

To derive the expression for EE,
the sum rate expression needs to be first derived.
To this end, the numbers of the serving SBSs in clusters $\mathcal{N}_{1}$ and $\mathcal{N}_{2}$ remain to be determined.
Under Scheme II, each SBS randomly selects BL and EL contents to cache in its local storage
under probability distributions $\mathbf{T}_{1}$ and $\mathbf{T}_{2}$, respectively.
As a result, the number of the serving SBSs follows the binomial distribution.
To be more specific, the number of serving SBSs in cluster $\mathcal{N}_{1}$
which cache BL content of the $f$-th file, i.e., $|\mathcal{N}_{1,f}|$,
follows the binomial distribution with parameters $N_{1}$ and $T_{1,f}$,
while $|\mathcal{N}_{2,f}|$ follows the binomial distribution with parameters $N_{2}$ and $T_{2,f}$.
Therefore, the sum rate of the designated user can be written as
\begin{align}
 & R_{\rm{Sum,2}}=\sum_{f=1}^{F}p_{f}\nonumber\\
 &(\underbrace{(1-T_{1,f})^{N_{1}}R_{M}(\gamma_{BL})+g_{\rm{HDV}}(f)(1-T_{2,f})^{N_{2}}R_{M}(\gamma_{EL})}_{\mathrm{served\:by\,the\,nearest\,MBS\,to\,obtain\,BL\,and\,EL\,contents}}\nonumber \\
 & +\sum_{n_{1}=1}^{N_{1}}\underbrace{C_{N_{1}}^{n_{1}}(T_{1,f})^{n_{1}}(1-T_{1,f})^{N_{1}-n_{1}}R_{S,BL}(\gamma_{BL},n_{1})}_{\mathrm{served\:by\,cooperative\,SBSs\,to\,obtain\,BL\,contents}}+\nonumber \\
 & \underbrace{g_{\rm{HDV}}(f)\sum_{n_{2}=1}^{N_{2}}C_{N_{2}}^{n_{2}}(T_{2,f})^{n_{2}}(1-T_{2,f})^{N_{2}-n_{2}}R_{S,EL}(\gamma_{EL},n_{2})}_{\mathrm{served\:by\,cooperative\,SBSs\,to\,obtain\,EL\,contents}}).\label{sr_s2}
\end{align}

\noindent From \eqref{sr_s2}, it can be concluded that
$R_{\rm sum,2}$ depends on the caching probabilities,
the number of serving SBSs, and the minimum QoS requirements.
Particularly, when there are more serving SBSs,
the requested video layers are more likely to be obtained locally.
This can lead to less power consumption
and relieve traffic congestion in backhaul.
According to \eqref{pt_s2} and \eqref{sr_s2},
the EE maximization problem with BL and EL caching probability design can be formulated as
\begin{subequations}\label{max_original-1}
\begin{align}
\underset{\mathbf{T}_{1},\mathbf{T}_{2}}{\mathrm{max}}\quad\: & EE_{2}=\frac{R_{\rm{Sum,2}}}{P_{\rm{Total,2}}}\label{eq:objective-1}\\
\mathrm{\mathrm{s.t.}}\:\quad\:\: & \sum_{f=1}^{F}T_{1,f}=M_{B},\label{eq:CacheSizeConstraints_1-1}\\
  & \sum_{f=1}^{F}T_{2,f}=M_{E},\label{eq:CacheSizeConstraints_2-1}\\
 & 0\leq T_{1,f}\leq1,\,0\leq T_{2,f}\leq1,\forall f,\label{eq:range-1}
\end{align}
\end{subequations}
\noindent where
(\ref{eq:CacheSizeConstraints_1-1}) and (\ref{eq:CacheSizeConstraints_2-1})
are the cache size constraints of each SBS when caching the BL and EL contents, respectively.
(\ref{eq:range-1}) specifies the feasible solution regions of the caching probabilities $T_{1,f}$ and $T_{2,f}$.
It is obvious that (\ref{max_original-1}) has the same form as (\ref{max_original}).
Therefore, in the same manner,
this problem can also be effectively solved by using the standard gradient projection method,
and more details can refer to Algorithm 1.

\begin{remark}
For Scheme II, the SBSs are capable of caching the complete video layers,
while Scheme I aims at caching parts of the BL and EL contents.
In Scheme II, video layers of the popular files are more likely to be completely cached
and deliveries from the backhaul links are prevented when these videos are requested.
In contrast, for Scheme I, each SBS caches parts of the popular video layers,
and the remaining contents need to be retrieved from backhauls,
consuming more power than caching.
As a consequence, it can be concluded that randomly caching the complete video layers
provides better EE than fractionally caching parts of them.
The superiority of Scheme II is demonstrated in terms of EE, especially in the presence of high backhaul power consumption.
These conclusions can be collaborated by extensive simulations as will be shown in Section VI.
\end{remark}

\section{Simulation Results}

\begin{table}[t]
\centering{}

\caption{Values of Simulation Parameters}

\begin{tabular}{c|c}
\hline
Parameter & Value\tabularnewline
\hline
$N_{1}$, $N_{2}$ & $4$\tabularnewline
\hline
$P_{S}$, $P_{M}$ & $23$ dBm (default), $43$ dBm\tabularnewline
\hline
$\lambda_{S},\lambda_{M}$ & $1/(100^{2}\pi)$, $1/(250^{2}\pi)$\tabularnewline
\hline
$a,b$ & 50 m, 100 m\tabularnewline
\hline
$\alpha_{S}$, $\alpha_{M}$ & $4$\tabularnewline
\hline
$M$ & $500$ Mbits (default)\tabularnewline
\hline
$F$ & 20\tabularnewline
\hline
$L_{B}$, $L_{E}$ & 100 Mbits, 200 Mbits \tabularnewline
\hline
$\alpha$ & $1$ (default)\tabularnewline
\hline
$W$ & $10$ Mbits\tabularnewline
\hline
$\gamma_{BL}$, $\gamma_{EL}$ & $10$ dB (default), $5$ dB (default)\tabularnewline
\hline
$\zeta_{S}$, $\zeta_{M}$ & 4.7\tabularnewline
\hline
$c_{ca}$ & $6.25\times10^{-12}$ W/bit \cite{Liu2015Energy}\tabularnewline
\hline
$c_{bh}$ & $5\times10^{-7}$ W/bit \cite{Liu2015Energy}\tabularnewline
\hline
$P_{\rm{S,Fix}}$ & $6.8$ W \cite{Yong2013Energy}\tabularnewline
\hline
$P_{\rm{M,Fix}}$ & $130 $ W\cite{Yong2013Energy}\tabularnewline
\hline
\end{tabular}
\end{table}

\begin{figure*}[t]
\centering{}\subfloat[$P(\mathrm{SIR}_{M}\geq\gamma)$ versus varying $\gamma$ under\protect \\
 different $P_{S}$.]{\includegraphics[scale=0.39]{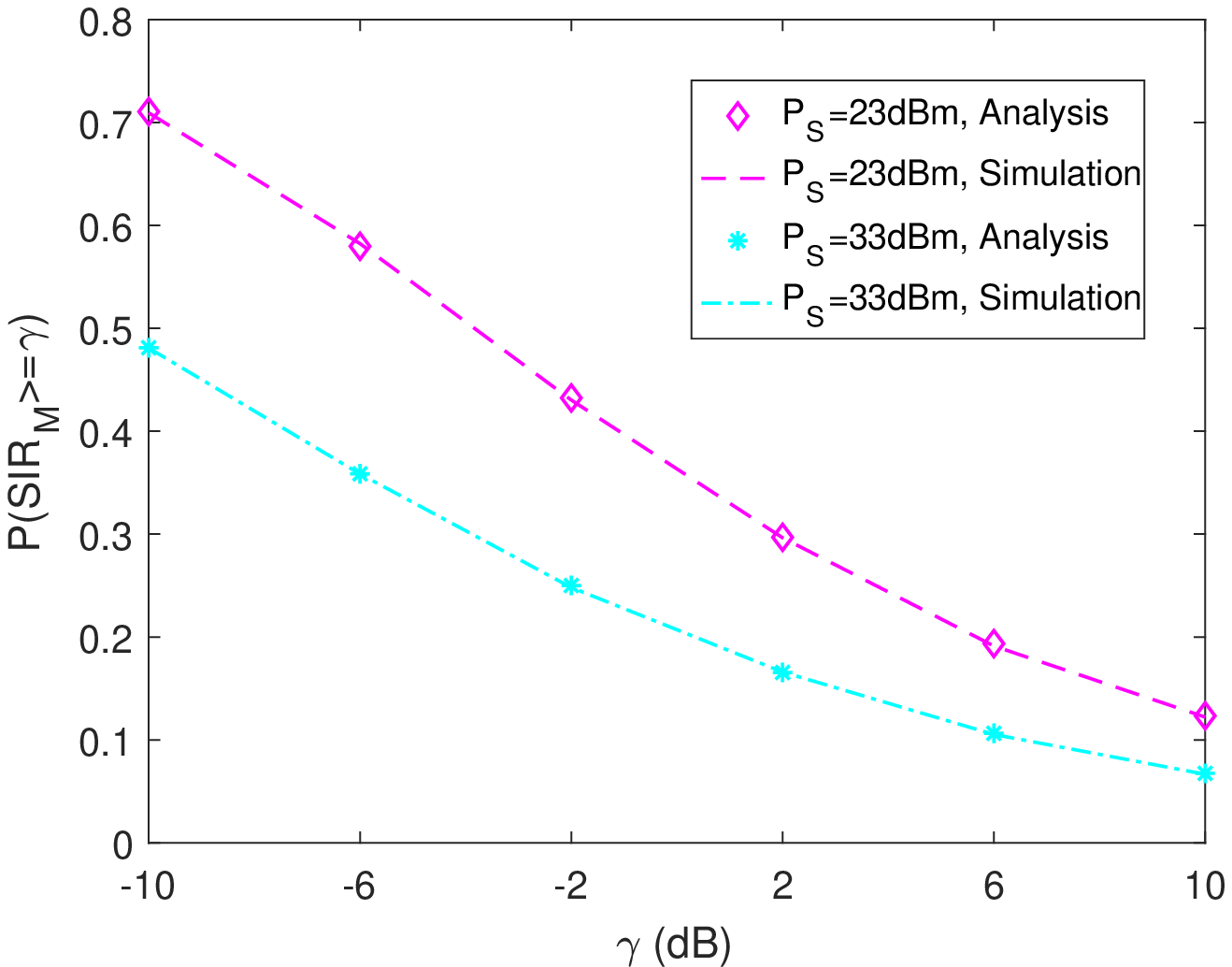}

}\subfloat[$P(\mathrm{SIR}_{S,BL}\geq\gamma_{BL})$ versus varying $\gamma_{BL}$\protect \\
   under different $n_{1}$.]{\includegraphics[scale=0.39]{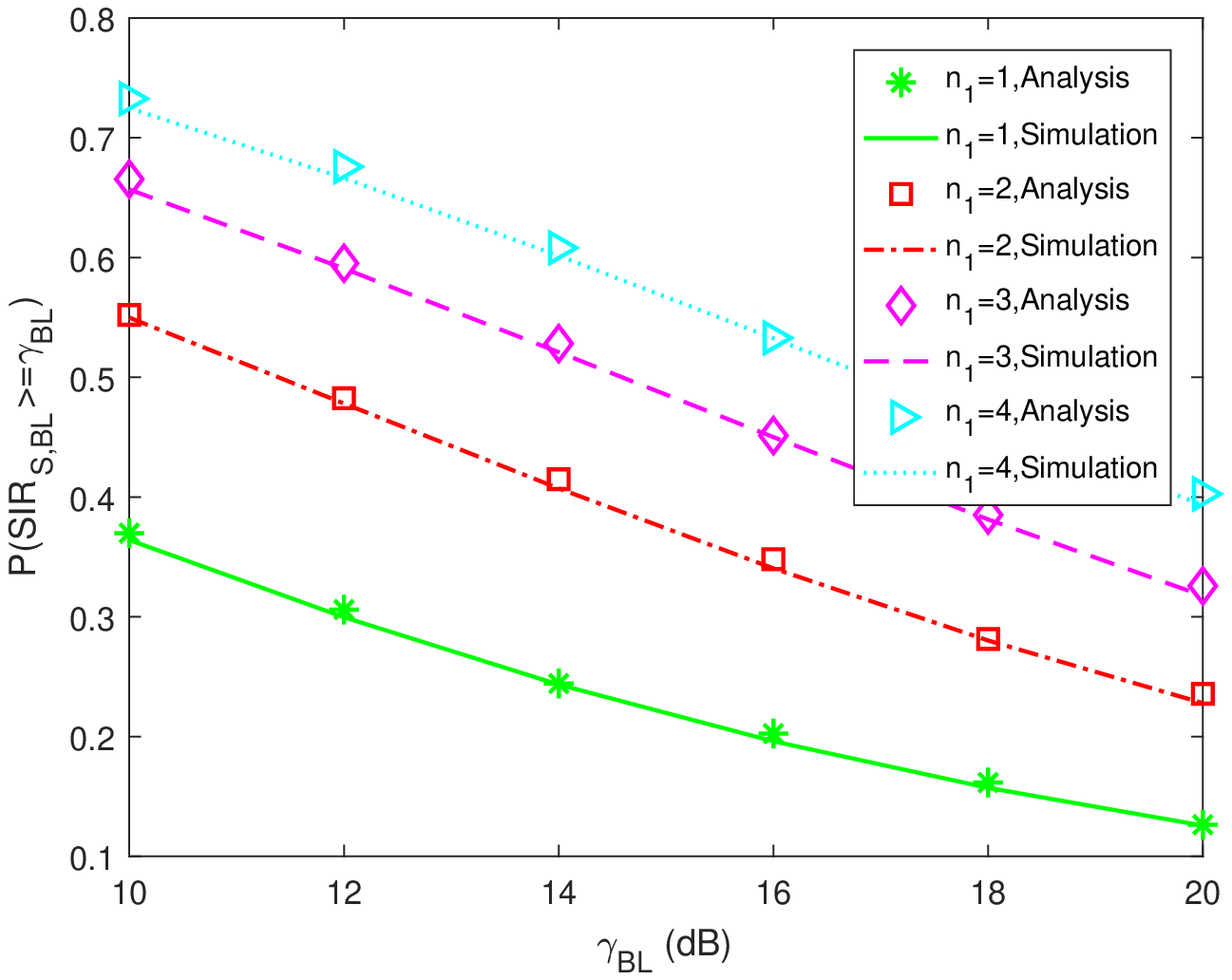}

}\subfloat[$P(\mathrm{SIR}_{S,EL}\geq\gamma_{EL})$ versus varying  $\gamma_{EL}$\protect \\
 under different $n_{2}$.]{\includegraphics[scale=0.39]{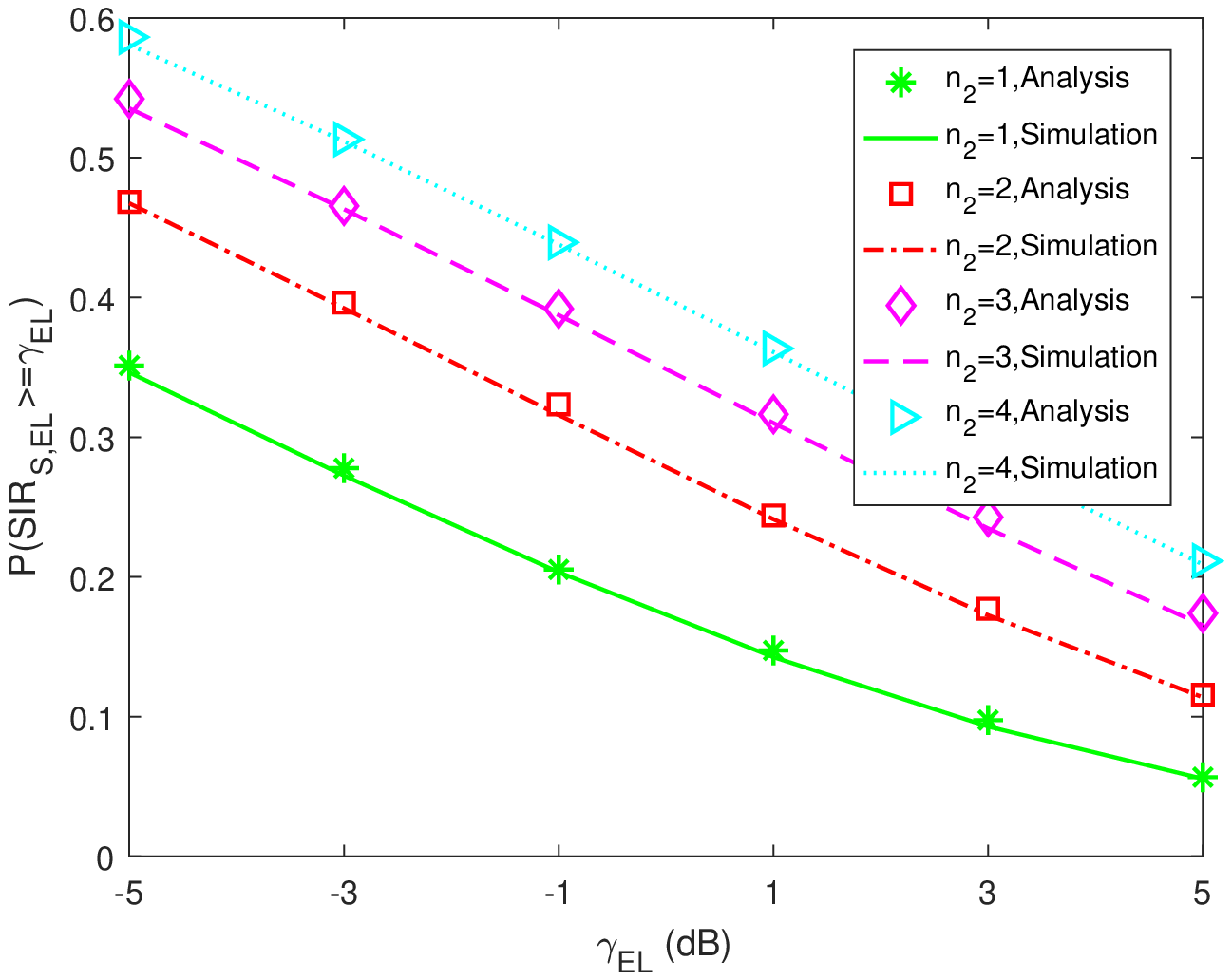}

}

\caption{The successful transmission probabilities when the user is served by the nearest MBS and cooperative SBSs.}
\end{figure*}

\begin{figure*}[t]
\centering{}\subfloat[$R_{M}(\gamma)$ versus varying $\gamma_{BL}$ under different \protect \\
$P_{S}$.]
{\includegraphics[scale=0.39]{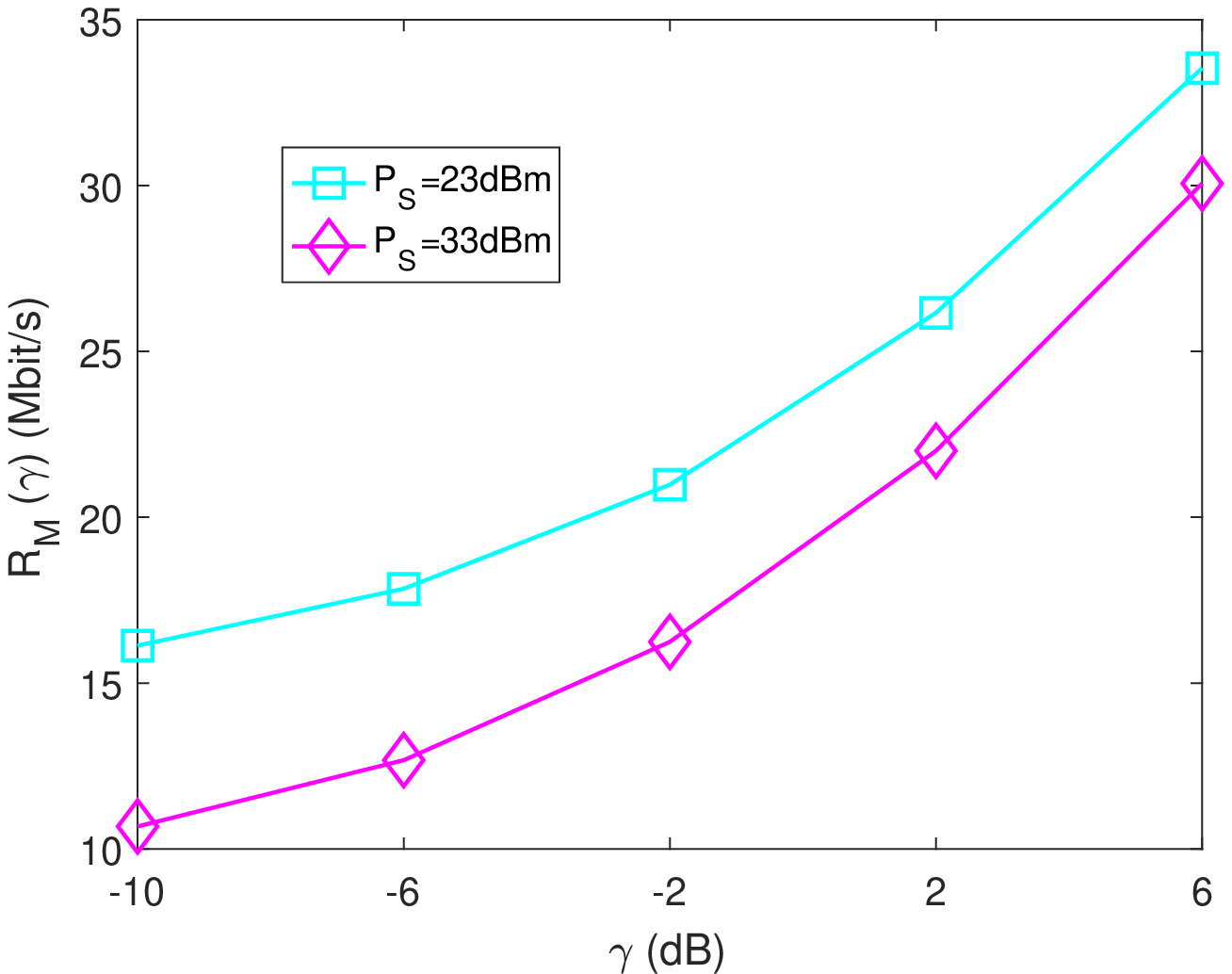}

}\subfloat[$R_{S,BL}(\gamma_{BL,}n_{1})$ versus varying $\gamma_{BL}$ \protect \\
 under different $n_{1}$.]{\includegraphics[scale=0.39]{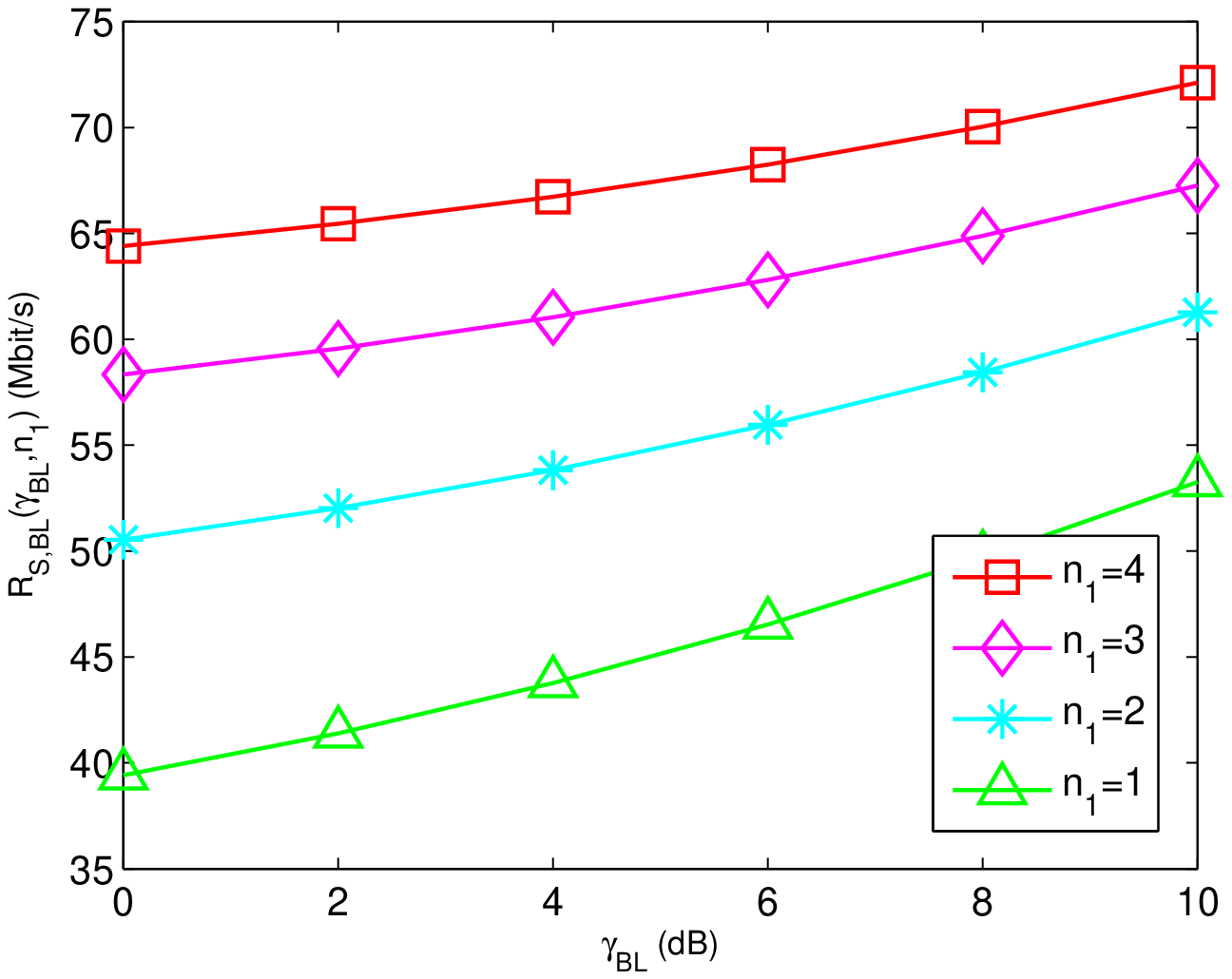}

}\subfloat[$R_{S,EL}(\gamma_{EL,}n_{2})$ versus varying $\gamma_{EL}$\protect \\
  under different $n_{2}$.]{\includegraphics[scale=0.39]{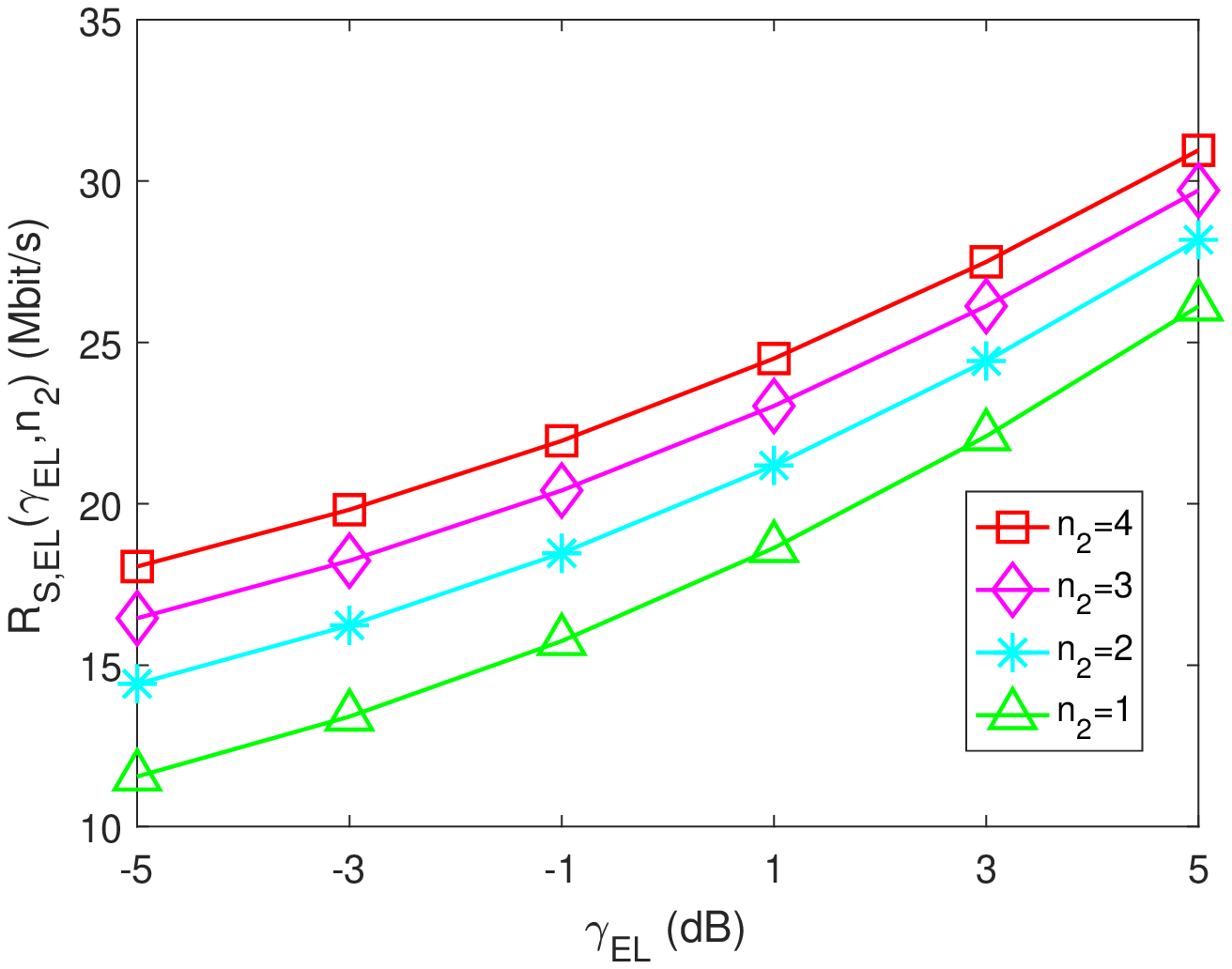}
}

\caption{The ergodic service rates when the user is served by the nearest MBS and cooperative SBSs.}
\end{figure*}

In this section, we show the simulation results of the
derived expressions for successful transmission probabilities and ergodic service rates,
as well as the EE performance under the proposed SVC-based caching schemes.
The simulation parameters are listed in Table I.
For comparison purpose,
three benchmarks are simulated, which are described as follows:

\begin{itemize}
\item Most Popular Content Placement (MPCP):
For SBSs located in clusters $\mathcal{N}_{1}$ and $\mathcal{N}_{2}$, $M_{B}$ BL contents
and $M_{E}$ EL contents of the most popular video files are cached in their local storage, respectively.

\item Uniform Content Placement (UCP): Regardless of the video popularity,
for SBSs located in clusters $\mathcal{N}_{1}$ and $\mathcal{N}_{2}$,
the same fractions of the BL and EL contents are cached, respectively.

\item Independent Content Placement (ICP):
The SBSs belonging to clusters $\mathcal{N}_{1}$ and $\mathcal{N}_{2}$ randomly
select $M_{B}$ and $M_{E}$ different BL and EL contents to cache in their local storage.
\end{itemize}

In Fig. 2, we show the successful transmission probabilities
derived from theoretical analysis and Monte Carlo simulations.
The plots of theoretical analysis and Monte Carlo simulations are overlapped,
confirming the accuracy of successful transmission probabilities given in Lemmas 1 and 2.
A common trend is also revealed that higher QoS requirements lead to lower successful transmission probabilities.
From Fig. 2 (a), it can be seen that a larger value of $P_{S}$ results in lower $P(\mathrm{SIR}_{M}\geq\gamma)$.
The reason for this is because, with the increase of $P_{S}$,
the co-existing SBSs can produce stronger interference towards the serving MBS,
and hence reduce $P(\mathrm{SIR}_{M}\geq\gamma)$.
Moreover, Figs. 2 (b) and (c) lead to the conclusion that with more cooperative SBSs,
the successful transmission probability can be improved.
\begin{figure}[t]
\centering{}\includegraphics[scale=0.55]{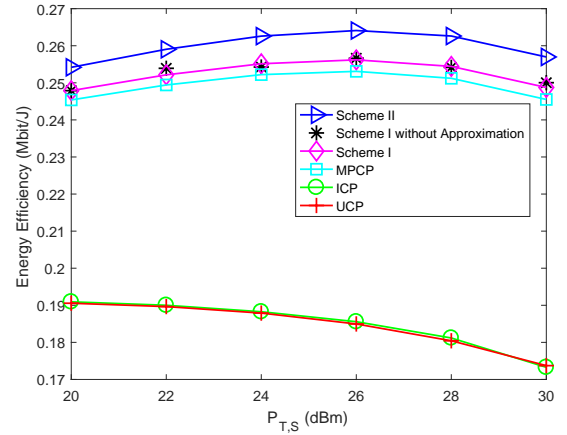}
\caption{The EE performance versus $P_{S}$ under different caching schemes.}
\end{figure}

In Fig. 3, we show the ergodic service rates,
i.e., $R_{M}(\gamma)$, $R_{S,BL}(\gamma_{BL},n_{1})$ and $R_{S,EL}(\gamma_{EL},n_{2})$,
with varying numbers of cooperative SBSs and QoS requirements.
Under our parameter settings,
the ergodic service rates increase as the QoS requirements grow,
since the improvement of the first term in (\ref{theo1}), (\ref{eq:Rate_S_BL_1}) and (\ref{eq:Rate_S_EL_1})
can make up for the loss of the second term.
However, higher $P_{S}$ can lead to lower $R_{M}(\gamma)$
due to a reduction of the successful transmission probability,
which coincides with the conclusion drawn from Fig. 2 (a).
Moreover, in Figs. 3 (b) and 3 (c),
the ergodic service rates are improved as $n_1$ and $n_2$ grow,
since there are more cooperative SBSs to enhance the successful transmission probabilities.
Therefore, it can be concluded that, in large scale heterogeneous networks with multiple SBSs,
the performance of ergodic service rates can be further enhanced.
\begin{figure}[t]
\centering{}\includegraphics[scale=0.55]{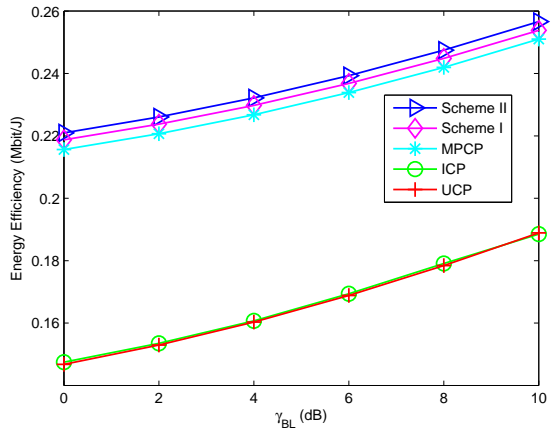}
\caption{The EE performance versus $\gamma_{BL}$ under different caching schemes.}
\end{figure}

Fig. 4 presents the relationship between the EE performance
and the transmit power of the SBSs under different caching schemes.
The superiority of our proposed SVC-based caching schemes is validated.
With the increase of $P_{S}$, the EE grows and the growth slows down.
When $P_{S}$ grows further, though the sum rate increases,
the improvement of the sum rate cannot scale up with the increase of the total power consumption.
This results in the degraded EE performance.
When we obtain the optimal caching fractions under Scheme I,
the optimal caching fractions are used to validate the accuracy of the $l_{0}$-norm approximation.
As shown in Fig. 4,
the approximation is accurate,
and the performance loss resulting from the approximation is marginal.
It can be concluded that when selecting the proper smooth parameter $\theta$,
the $l_{0}$-norm of the caching fractions can be accurately estimated.
For the ICP scheme, the file selection process is performed under each channel realization.
When there are sufficient channel realizations,
the EE of ICP is comparable to that of UCP.

\begin{figure}[t]
\centering{}\includegraphics[scale=0.55]{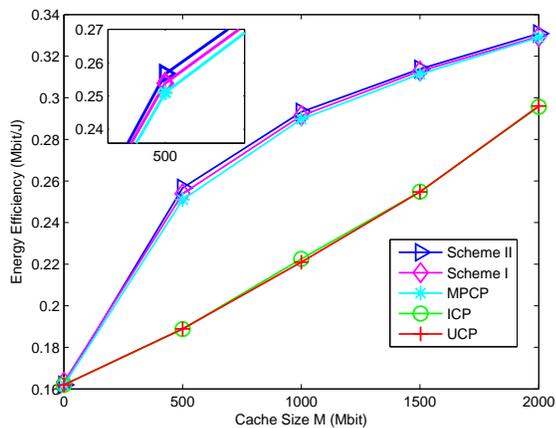}
\caption{The EE performance versus cache size under different caching schemes.}
\end{figure}
Fig. 5 shows the EE performance of different caching schemes under various values of  $\gamma_{BL}$.
It is obvious that the proposed caching schemes are superior to the three benchmarks,
especially achieving much more performance gain than the UCP and ICP schemes.
Furthermore, it can be seen that Scheme II can achieve higher EE than Scheme I.
This indicates that randomly caching complete video layers
provides better EE than storing only parts of them,
even under the optimal caching fractions.

In Fig. 6, we present the EE performance of different caching schemes with varying cache size $M$.
When a larger cache size is equipped at each SBS,
more video contents can be locally cached.
Therefore, the demand for backhaul links can be relieved substantially,
which in turn can significantly reduce service delay and backhaul power consumption.
In practice, for video files with the same sizes,
caching them can consume less power consumption than retrieving them from microwave backhaul links.
As a result, larger cache sizes can lead to higher EE.
Particularly, when $M=0$, all caching schemes have the same EE.
This is equivalent to the case with no caching.
The proposed caching schemes provide better EE in small cache size region,
and reach the maximum performance gap at about $M=600$ Mbits.
However, when the cache size grows further,
the performance gaps gradually diminish,
since all video files will be cached when the cache size of each SBS is large enough.

\begin{figure}[t]
\centering{}\includegraphics[scale=0.55]{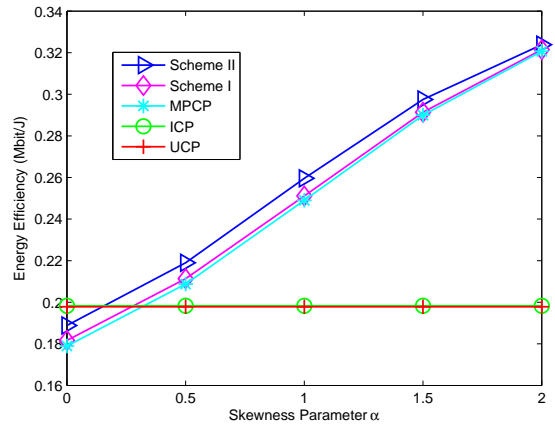}
\caption{The EE performance versus skewness parameter under different caching schemes.}
\end{figure}

In Fig. 7, the relationship between the EE performance and skewness parameter $\alpha$ is plotted.
Note that larger $\alpha$ means that fewer video files can meet the majority of user requests.
Therefore, video files with smaller indices are much more likely to be stored in the local cache of SBSs.
When $\alpha$ is small, the UCP and ICP schemes provide better EE.
In this case, video popularities are uniformly distributed,
and the UCP and ICP schemes are able to increase file diversity, and then increase request hit ratio.
\begin{figure}[t]
\centering{}\includegraphics[scale=0.55]{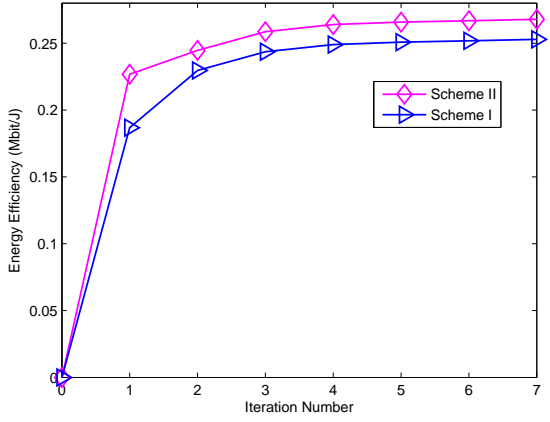}
\caption{The convergence property of the proposed algorithm.}
\end{figure}

In Fig. 8, we show the convergence property of the proposed algorithm.
It can be seen that the proposed algorithm is able to converge
after a small number of iterations,
which validates the effectiveness of the standard gradient projection method in practical implementations.
\section{Conclusion}
This paper proposed two energy-efficient SVC-based caching schemes
to boost EE in cache-enabled heterogeneous networks.
Based on the proposed caching schemes,
we established the power consumption model,
and derived the expressions for successful transmission probabilities and ergodic service rates.
We further formulated two EE maximization problems,
which are subject to the cache size constraint of each SBS.
After taking approximations of the $l_{0}$-norm,
the EE optimization problems can be efficiently solved.
Numerical results confirmed the accuracy of our analysis
as well as the superiority of the proposed caching schemes to three benchmarks.

\begin{appendices}
\section{Proof of Lemma 1}
In the proposed model, the designated user is located at the origin of the observed network
and the distance between the nearest MBS and the user is denoted by $r_{M,m_{0}}$,
whose probability density function (PDF) is shown as\cite{Chen2017Cooperative123}
\begin{gather}
f_{m_{0}}(r_{M,m_{0}})=2\pi\lambda_{M}r_{M,m_{0}}e^{-\lambda_{M}\pi r_{M,m_{0}}^{2}}.\label{eq:PDF_1}
\end{gather}
\noindent Then, $P(\mathrm{SIR}_{M}\geq\gamma)$ is calculated as
\begin{gather}
P(\mathrm{SIR}_{M}\geq\gamma)=\int_{0}^{\infty}f_{m_{0}}(x)P(\mathrm{SIR}_{M}\geq\gamma|r_{M,m_{0}}=x)\mathrm{d}x.\label{eq:145}
\end{gather}
\noindent For simplicity of notations, the interference from all SBSs and other non-serving MBSs are denoted by
\noindent
\begin{align*}
&I_{S_{1}}=\sum_{n\in\mathrm{\Phi_{S}}\setminus\mathcal{N}_{2}}\left|h_{S,n}\right|^{2}P_{S}r_{S,n}^{-\alpha_{S}},\\
&I_{M_{1}}=\sum_{m\in\mathrm{\Phi_{M}\setminus}m_{0}}\left|h_{M,m}\right|^{2}P_{M}r_{M,m}^{-\alpha_{M}}.
\end{align*}
\noindent Relying on stochastic geometry,
$P(\mathrm{SIR}_{M}\geq\gamma|r_{M,m_{0}}=x)$ can be calculated as
\begin{align}
 &P(\mathrm{SIR}_{M}\geq\gamma|r_{M,m_{0}}=x)\nonumber\\
 &=P(\frac{\left|h_{M,m_{0}}\right|^{2}P_{M}r_{M,m_{0}}^{-\alpha_{M}}}{I_{S_{1}}+I_{M_{1}}}\geq\gamma|r_{M,m_{0}}=x)\nonumber\\
 &=P(\left|h_{M,m_{0}}\right|^{2}\geq\gamma P_{M}^{-1}x^{\alpha_{M}}(I_{S_{1}}+I_{M_{1}}))\nonumber\\
 &\stackrel{(a)}{=}\mathbb{E}_{I_{S_{1}},I_{M_{1}}}[\exp(-(I_{S_{1}}+I_{M_{1}})\gamma P_{M}^{-1}x^{\alpha_{M}})]\nonumber\\
 &=\mathcal{L}_{I_{S_{1}}}(\gamma P_{M}^{-1}x^{\alpha_{M}})\mathcal{L}_{I_{M_{1}}}(\gamma P_{M}^{-1}x^{\alpha_{M}}),\label{eq:we}
 \end{align}
\noindent where $(a)$ follows the fact that $\left|h_{M,m_{0}}\right|^{2}\sim\exp(1)$
and $\exp(\mu)$ denotes the exponential distribution with mean $\mu$.
Additionally, $\mathcal{L}_{I_{S_{1}}}(\gamma P_{M}^{-1}x^{\alpha_{M}})$ and $\mathcal{L}_{I_{M_{1}}}(\gamma P_{M}^{-1}x^{\alpha_{M}})$
are the Laplace transforms of interference $I_{S_{1}}$ and $I_{M_{1}}$, respectively.
Let $k_{1}=\gamma P_{M}^{-1}x^{\alpha_{M}}$.
Then, $\mathcal{L}_{I_{S_{1}}}(k_{1})$ can be obtained as follows
\begin{align}
 &\mathcal{L}_{I_{S_{1}}}(k_{1})=\mathbb{E}_{I_{S_{1}}}[\exp(-\sum_{n\in\mathrm{\Phi_{S}}}k_{1}I_{S_{1},n})]\nonumber\\
 &=\mathbb{E}_{I_{S_{1}}}[\prod_{n\in\Phi_{S}}\exp(-k_{1}\left|h_{S,n}\right|^{2}P_{S}r_{S,n}^{-\alpha_{S}})]\nonumber \\
 & =\mathbb{E}_{I_{S_{1}}}[\prod_{n\in\Phi_{S}}\frac{1}{1+k_{1}P_{S}r_{S,n}^{-\alpha_{S}}}]\nonumber\\
 &=\exp(-2\pi\lambda_{s}\int_{0}^{\infty}(1-\frac{1}{1+k_{1}P_{S}\rho^{-\alpha_{S}}})\rho\mathrm{d}\rho\nonumber \\
 & =\exp(-\pi\lambda_{s}(k_{1}P_{S})^{\frac{2}{\alpha_{S}}}\int_{0}^{\infty}\frac{1}{1+t^{\frac{\alpha_{S}}{2}}}\mathrm{d}t).\label{eq:L1}
\end{align}
\noindent We denote $G_{\alpha}(x)=\int_{x}^{\infty}\frac{1}{1+t^{\frac{\alpha}{2}}}\mathrm{d}t$,
then $\mathcal{L}_{I_{S_{1}}}(k_{1})=\exp(-\pi\lambda_{s}(k_{1}P_{S})^{\frac{2}{\alpha_{S}}}G_{\alpha_{S}}(0))$.
In the similar manner, $\mathcal{L}_{I_{M_{1}}}(k_{1})$ can be calculated as
\begin{align}
 &\mathcal{L}_{I_{M_{1}}}(k_{1})=\mathbb{E}_{I_{M_{1}}}[\exp(-\sum_{m\in\mathrm{\Phi_{M}\setminus}m_{0}}k_{1}I_{M_{1},m})]\nonumber\\
 &=\mathbb{E}_{I_{M_{1}}}[\prod_{m\in\mathrm{\Phi_{M}\setminus}m_{0}}\exp(-k_{1}\left|h_{M,m}\right|^{2}P_{M}r_{M,m}^{-\alpha_{M}})]\nonumber \\
 & =\mathbb{E}_{I_{M_{1}}}[\prod_{m\in\mathrm{\Phi_{M}\setminus}m_{0}}\frac{1}{1+k_{1}P_{M}r_{M,m}^{-\alpha_{M}})}\nonumber\\
 &=\exp(-2\pi\lambda_{M}\int_{x}^{\infty}(1-\frac{1}{1+k_{1}P_{M}\rho^{-\alpha_{M}}})\rho\mathrm{d}\rho\nonumber \\
 %& =\exp(-\pi\lambda_{M}(k_{1}P_{M})^{\frac{2}{\alpha_{M}}}\int_{x^{2}(k_{1}P_{M})^{-\frac{2}{\alpha_{M}}}}^{\infty}\frac{1}{1+t^{\frac{\alpha_{M}}{2}}}\mathrm{d}t)\nonumber
 & =\exp(-\pi\lambda_{M}(k_{1}P_{M})^{\frac{2}{\alpha_{M}}}G_{\alpha_{M}}(x^{2}(k_{1}P_{M})^{-\frac{2}{\alpha_{M}}})).\label{eq:L2}
\end{align}
Therefore, we can obtain that
\begin{align}
 &P(\mathrm{SIR}_{M}\geq\gamma|r_{M,m_{0}}=x)=\exp(-\pi\lambda_{M}x^{2}\gamma{}^{\frac{2}{\alpha_{M}}}G_{\alpha_{M}}(\gamma{}^{-\frac{2}{\alpha_{M}}}))\nonumber\\ &\exp(-\pi\lambda_{S}(\gamma\frac{P_{S}}{P_{M}}x^{\alpha_{M}})^{\frac{2}{\alpha_{S}}}G_{\alpha_{S}}(0)).\label{eq:vbyu}
\end{align}
\noindent Finally, substituting (\ref{eq:vbyu}) and (\ref{eq:PDF_1}) into (\ref{eq:145}), $P(\mathrm{SIR}_{M}\geq\gamma)$ is obtained.
\qed
\section{Proof of Corollary 1}
When $\alpha_{M}=\alpha_{S}=4$, it is intuitional to obtain that $G_{\alpha_{S}}(0)=\frac{\pi}{2}$
and $G_{\alpha_{M}}(\gamma{}^{-\frac{2}{\alpha_{M}}})=\frac{\pi}{2}-\mathrm{arctan}(\gamma{}^{-\frac{2}{\alpha_{M}}})=\mathrm{arccot}(\gamma{}^{-\frac{2}{\alpha_{M}}})$.
Next, the simplified form of $P(\mathrm{SIR}_{M}\geq\gamma)$ can be obtained as follows
\begin{align*}
 &P(\mathrm{SIR}_{M}\geq\gamma)=\int_{0}^{\infty}f_{m_{0}}(x)\exp(-\pi x^{2}\gamma{}^{\frac{1}{2}}(\frac{\pi}{2}\lambda_{S}(\frac{P_{S}}{P_{M}})^{\frac{1}{2}}\\
 &+\lambda_{M}\mathrm{arccot}(\gamma{}^{-\frac{1}{2}})))\mathrm{d}x\\
  & =\pi\lambda_{M}\int_{0}^{\infty}\exp(-\pi x^{2}(\lambda_{M}\\
 &+\gamma^{\frac{1}{2}}(\frac{\pi}{2}\lambda_{S}(\frac{P_{S}}{P_{M}})^{\frac{1}{2}}+\lambda_{M}\mathrm{arccot}(\gamma{}^{-\frac{1}{2}}))))\mathrm{d}x^{2}\\
 \end{align*}
 \begin{align*}
 &=\lambda_{M}(\lambda_{M}+\gamma{}^{\frac{1}{2}}(\frac{\pi}{2}\lambda_{S}(\frac{P_{S}}{P_{M}})^{\frac{1}{2}}+\lambda_{M}\mathrm{arccot}(\gamma{}^{-\frac{1}{2}})))^{-1}\\
 &=(1+\lambda_{M}^{-1}\gamma{}^{\frac{1}{2}}(\frac{\pi}{2}\lambda_{S}(\frac{P_{S}}{P_{M}})^{\frac{1}{2}}+\lambda_{M}\mathrm{arccot}(\gamma{}^{-\frac{1}{2}})))^{-1}.\qed
\end{align*}

\section{Proof of Theorem 1}
For notational simplicity, let $y_{M}=\mathrm{SIR}_{M}$ and $\Omega(y_{M})$ denotes the successful transmission event $\mathrm{SIR}_{M}\geq\gamma$.
The conditional PDF of $y_{M}$ is then denoted as $g_(y_{M}|\Omega(y_{M}))$.
Therefore, we have the following derivations
\begin{align}
 & \mathbb{E}[\log_{2}(1+y_{M})|\Omega(y_{M})]\nonumber \\
 &=\int_{0}^{\infty}f_{m_{0}}(x)\mathrm{d}x\int_{0}^{\infty}\log_{2}(1+y_{M})g(y_{M}|\Omega(y_{M}))\mathrm{d}y_{M}\nonumber \\
 & =\frac{1}{\ln2}\int_{0}^{\infty}f_{m_{0}}(x)\mathrm{d}x\int_{0}^{\infty}\ln(1+y_{M})g(y_{M}|\Omega(y_{M}))\mathrm{d}y_{M}\nonumber\\
 & =\frac{1}{\ln2}\int_{0}^{\infty}f_{m_{0}}(x)\mathrm{d}x\int_{0}^{\infty}(\int_{0}^{y_{M}}\frac{1}{1+t}dt)g(y_{M}|\Omega(y_{M})\mathrm{d}y_{M}\nonumber\\
 & \stackrel{(b)}{=}\frac{1}{\ln2}\int_{0}^{\infty}f_{m_{0}}(x)\mathrm{d}x\int_{0}^{\infty}\frac{1}{1+t}\mathrm{d}t\int_{t}^{\infty}g(y_{M}|\Omega(y_{M}))\mathrm{d}y_{M}\nonumber\\
 & \stackrel{(c)}{=}\frac{1}{\ln2}\int_{0}^{\infty}f_{m_{0}}(x)\mathrm{d}x\int_{0}^{\infty}\frac{1}{1+t}P(y_{M}\geq t|\Omega(y_{M}))\mathrm{d}t\nonumber\\
 & =\frac{1}{\ln2}\int_{0}^{\infty}f_{m_{0}}(x)\mathrm{d}x\nonumber\\
 &\int_{0}^{\infty}\frac{1}{1+t}\frac{P(\mathrm{SIR}_{M}\geq t,\mathrm{SIR}_{M}\geq\gamma|r_{M,m_{0}}=x)}{P(\mathrm{SIR}_{M}\geq\gamma|r_{M,m_{0}}=x)}\mathrm{d}t\nonumber\\
 & =\frac{1}{\ln2}\int_{0}^{\infty}f_{m_{0}}(x)\mathrm{d}x\nonumber\\
 &\int_{0}^{\infty}\frac{1}{1+t}\frac{P(\mathrm{SIR}_{M}\geq \rm{max}(t,\gamma)|r_{M,m_{0}}=x)}{P(\mathrm{SIR}_{M}\geq\gamma|r_{M,m_{0}}=x)}\mathrm{d}t\nonumber\\
 & =\frac{1}{\ln2}\int_{0}^{\infty}f_{m_{0}}(x)\mathrm{d}x\nonumber\\
 &\left[\int_{0}^{\gamma}\frac{1}{1+t}\mathrm{d}t+\int_{\gamma}^{\infty}\frac{P(\mathrm{SIR}_{M}\geq t|r_{M,m_{0}}=x)}{P(\mathrm{SIR}_{M}\geq\gamma|r_{M,m_{0}}=x)(1+t)}\mathrm{d}t\right]\nonumber\\
 & =\log_{2}(1+\gamma)+\frac{1}{\ln2}\nonumber\\
 &\int_{0}^{\infty}f_{m_{0}}(x)\mathrm{d}x\int_{\gamma}^{\infty}\frac{P(\mathrm{SIR}_{M}\geq t|r_{M,m_{0}}=x)}{P(\mathrm{SIR}_{M}\geq\gamma|r_{M,m_{0}}=x)(1+t)}\mathrm{d}t,\label{eq:lgu}
\end{align}
\noindent where $(b)$ and $(c)$ follow the facts that
\begin{align*}
&\int_{0}^{\infty}(\int_{0}^{y_{M}}\frac{1}{1+t}dt)g(y_{M}|\Omega(y_{M})\mathrm{d}y_{M}\nonumber\\
&=\int_{0}^{\infty}\frac{1}{1+t}\mathrm{d}t\int_{t}^{\infty}g(y_{M}|\Omega(y_{M}))\mathrm{d}y_{M},\\
&\int_{t}^{\infty}g(y_{M}|\Omega(y_{M}))\mathrm{d}y_{M}=P(y_{M}\geq t|\Omega(y_{M})).
\end{align*}
\noindent Finally, substituting (\ref{eq:lgu}) and (\ref{eq:PDF_1}) into (\ref{eq:Service Rate_M}),
the expression for $R_{M}(\gamma)$ shown in (\ref{theo1}) is obtained.
\qed
\section{Proof of Lemma 2}
Let vectors $\mathbf{r}_{S_{BL}}=[r_{BL,1},r_{BL,2},...,r_{BL,n_{1}}]$ and $\mathbf{r}_{S_{EL}}=[r_{EL,1},r_{EL,2},...,r_{EL,n_{2}}]$
represent the positions of the serving SBSs located in two clusters.
The serving SBSs in clusters $\mathcal{N}_{1}$ and $\mathcal{N}_{2}$ are independently and uniformly distributed
within the circle with radius $a$ and annulus with radii $a$ and $b$.
Therefore, the joint PDF of $\mathbf{r}_{S_{BL}}$ and $\mathbf{r}_{S_{EL}}$ are given by \cite{Chen2017Cooperative123}
\begin{align}
f(r_{BL,1},...,r_{BL,n_{1}})=\stackrel[k=1]{n_{1}}{\prod}\frac{2r_{BL,k}}{a^{2}},\label{PDF_1}
\end{align}
\begin{align}
f(r_{EL,1},...,r_{EL,n_{2}})=\stackrel[k=1]{n_{2}}{\prod}\frac{2r_{EL,k}}{b^{2}-a^{2}}.\label{PDF_2}
\end{align}
\subsection{The derivation of $P(\mathrm{SIR}_{S,BL}\geq\gamma_{BL})$}
From the definition of successful transmission probability,
$P(\mathrm{SIR}_{S,BL}\geq\gamma_{BL})$ can be expressed as
\begin{align}
 P(\mathrm{SIR}_{S,BL}\geq&\gamma_{BL})=\int_{0}^{a}...\int_{0}^{a}f(x_{BL,1},...,x_{BL,n_{1}})\nonumber\\
 &P(\mathrm{SIR}_{S,BL}\geq\gamma_{BL}|\mathbf{r}_{S_{BL}}=\mathbf{x}_{BL})\mathrm{d}\mathbf{x}_{BL}.\label{eq:mnb}
\end{align}
For notational simplicity, we let
\noindent
\begin{align*}
&I_{S_{2}}=\sum_{n\in\mathrm{\Phi_{S}}\setminus\mathcal{N}_{1}}\left|h_{S,n}\right|^{2}P_{S}r_{S,n}^{-\alpha_{S}},\\
&I_{M_{2}}=\sum_{m\in\mathrm{\Phi_{M}}}\left|h_{M,m}\right|^{2}P_{M}r_{M,m}^{-\alpha_{M}}.
\end{align*}
\noindent Next, $P(\mathrm{SIR}_{S,BL}\geq\gamma_{BL}|\mathbf{r}_{S_{BL}}=\mathbf{x}_{BL})$ is calculated as
\begin{align}
&P(\mathrm{SIR}_{S,BL}\geq\gamma_{BL}|\mathbf{r}_{S_{BL}}=\mathbf{x}_{BL})\nonumber\\
&=P(\left|\sum_{k=1}^{n_{1}}h_{S,k}\sqrt{P_{S}}x_{BL,k}^{-\frac{\alpha_{S}}{2}}\right|^{2}\geq\gamma_{BL}(I_{S_{2}}+I_{M_{2}}))\nonumber\\
&\stackrel{(d)}{=}\mathbb{E}_{I_{S_{2}},I_{M_{2}}}[\exp(-\frac{1}{\sum_{k=1}^{n_{1}}x_{BL,k}^{-\alpha_{S}}}\gamma_{BL}(I_{S_{2}}+I_{M_{2}})P_{S}^{-1})]\nonumber\\
&=\mathcal{L}_{I_{S_{2}}}(\frac{\gamma_{BL}P_{S}^{-1}}{\sum_{k=1}^{n_{1}}x_{BL,k}^{-\alpha_{S}}})\mathcal{L}_{I_{M_{2}}}(\frac{\gamma_{BL}P_{S}^{-1}}{\sum_{k=1}^{n_{1}}x_{BL,k}^{-\alpha_{S}}}),\label{eq:vbhu}
\end{align}
\noindent where (d) follows the fact that $\left|\sum_{k=1}^{n_{1}}h_{S,k}\sqrt{P_{S}}r_{BL,k}^{-\frac{\alpha_{S}}{2}}\right|^{2}\sim P_{S}^{-1}\exp(\frac{1}{\sum_{k=1}^{n_{1}}x_{BL,k}^{-\alpha_{S}}}).$
Let $k_{2}=\frac{\gamma_{BL}P_{S}^{-1}}{\sum_{k=1}^{n_{1}}x_{BL,k}^{-\alpha_{S}}}$.
Following the similar steps shown before, $\mathcal{L}_{I_{S_{2}}}(k_{2})$ and $\mathcal{L}_{I_{M_{2}}}(k_{2})$ are calculated as
\begin{align}
\mathcal{L}_{I_{S_{2}}}(k_{2}) & =\exp(-\pi\lambda_{S}(k_{2}P_{S})^{\frac{2}{\alpha_{S}}}G_{\alpha_{S}}(a^{2}(k_{2}P_{S})^{-\frac{2}{\alpha_{S}}})),\label{eq:1256}
\end{align}
\begin{equation}
\mathcal{L}_{I_{M_{2}}}(k_{2})=\exp(-\pi\lambda_{M}(k_{2}P_{M})^{\frac{2}{\alpha_{M}}}G_{\alpha_{M}}(0)),\label{eq:1123}
\end{equation}
\noindent respectively. Substituting formulas (\ref{eq:1256}) and (\ref{eq:1123}) into (\ref{eq:vbhu}),
we can obtain that
\begin{align}
&P(\mathrm{SIR}_{S,BL}\geq\gamma_{BL}|\mathbf{r}_{S_{BL}}=\mathbf{x}_{BL})=\nonumber\\
&\exp(-\pi\lambda_{S}c{}^{\frac{2}{\alpha_{S}}}G_{\alpha_{S}}(a^{2}c^{-\frac{2}{\alpha_{S}}})-\pi\lambda_{M}(c\frac{P_{M}}{P_{S}})^{\frac{2}{\alpha_{M}}}G_{\alpha_{M}}(0)).\label{eq:cv}
\end{align}
Finally, substituting (\ref{eq:cv}) and (\ref{PDF_1}) into (\ref{eq:mnb}),
we can obtain the expression for $P(\mathrm{SIR}_{S,BL}\geq\gamma_{BL})$.
Thus, the proof of $P(\mathrm{SIR}_{S,BL}\geq\gamma_{BL})$ is completed.

\subsection{The derivation of $P(\mathrm{SIR}_{S,EL}\geq\gamma_{EL})$}
Following the similar methods, $P(\mathrm{SIR}_{S,EL}\geq\gamma_{EL})$ can be expressed as
\begin{align}
P(\mathrm{SIR}_{S,EL}\geq&\gamma_{EL})=\int_{a}^{b}...\int_{a}^{b} f(x_{EL,1},...,x_{EL,n_{2}})\nonumber\\
&P(\mathrm{SIR}_{S,EL}\geq\gamma_{EL}|\mathbf{r}_{S_{EL}}=\mathbf{x}_{EL})\mathrm{d}\mathbf{x}_{EL}.\label{kli}
\end{align}
\noindent In order to simplify notations, we let
\begin{gather*}
I_{S_{3}}=\sum_{n\in\mathcal{N}_{1}}\left|h_{S,n}\right|^{2}P_{S}r_{S,n}^{-\alpha_{S}},\\ I_{S_{4}}=\sum_{n\in\mathrm{\Phi_{S}}\setminus\{\mathcal{N}_{1}\cup\mathcal{N}_{2}\}}\left|h_{S,n}\right|^{2}P_{S}r_{S,n}^{-\alpha_{S}}.
\end{gather*}
\noindent Afterwards, the expression for $P(\mathrm{SIR}_{S,EL}\geq\gamma_{EL}|\mathbf{r}_{S_{EL}}=\mathbf{x}_{EL})$ can be calculated as
\begin{align}
P(\mathrm{SIR}_{S,EL}&\geq\gamma_{EL}|\mathbf{r}_{S_{EL}}=\mathbf{x}_{EL})=\nonumber\\
&\mathcal{L}_{I_{S_{3}}}(k_{3})\mathcal{L}_{I_{S_{4}}}(k_{3})\mathcal{L}_{I_{M_{2}}}(k_{3}),\label{cdf_el}
\end{align}
\noindent where
\begin{gather}
k_{3}=\frac{\gamma_{EL}P_{S}^{-1}}{\sum_{k=1}^{n_{2}}x_{EL,k}^{-\alpha_{S}}},\label{eq:1b11}
\end{gather}
\begin{align}
&\mathcal{L}_{I_{S_{3}}}(k_{3})=\exp(-\pi\lambda_{S}(k_{3}P_{S})^{\frac{2}{\alpha_{S}}}\int_{0}^{a^{2}(k_{3}P_{S})^{-\frac{2}{\alpha_{S}}}}\frac{1}{1+t^{\frac{\alpha_{S}}{2}}}\mathrm{d}t),\label{eq:1b}
\end{align}
\begin{gather}
\mathcal{L}_{I_{S_{4}}}(k_{3})=\exp(-\pi\lambda_{S}(k_{3}P_{S})^{\frac{2}{\alpha_{S}}}G_{\alpha_{S}}(b^{2}(k_{3}P_{S})^{-\frac{2}{\alpha_{S}}}).\label{eq:1c}
\end{gather}
\noindent Thus, (\ref{cdf_el}) can be rewritten as
\begin{align}
 &P(\mathrm{SIR}_{S,EL}\geq\gamma_{EL}|\mathbf{r}_{S_{EL}}=\mathbf{x}_{EL})=\nonumber\\
&\exp(-\pi(\lambda_{M}(d\frac{P_{M}}{P_{S}})^{\frac{2}{\alpha_{M}}}G_{\alpha_{M}}(0)\nonumber\\
&-\lambda_{S}d{}^{\frac{2}{\alpha_{S}}}(\int_{0}^{a^{2}d^{-\frac{2}{\alpha_{S}}}}\frac{1}{1+t^{\frac{\alpha_{S}}{2}}}\mathrm{d}t)+G_{\alpha_{S}}(b^{2}d^{-\frac{2}{\alpha_{S}}}))).\label{eq:bn}
\end{align}
Substituting (\ref{eq:bn}) and (\ref{PDF_2}) into (\ref{kli}),
we can obtain the theoretical expression for $P(\mathrm{SIR}_{S,EL}\geq\gamma_{EL})$.
\qed
\end{appendices}

\bibliographystyle{IEEEtran}
\bibliography{ciations}

\end{document}